\documentclass[aps, pra, notitlepage, citeautoscript, superscriptaddress, reprint]{revtex4-2}

\usepackage{xr}
\usepackage{amsmath}
\usepackage{amsfonts,amssymb,amsthm,amsxtra}
\usepackage[]{graphicx}
\usepackage{grffile}
\usepackage{mathrsfs}
\usepackage{framed}
\usepackage{bbm}
\usepackage{bm}
\usepackage{braket}
\usepackage{xcolor}
\usepackage{mathtools}
\usepackage[bookmarks=false,colorlinks=true,urlcolor=blue,citecolor=blue,linkcolor=blue]{hyperref}

\DeclareMathOperator*{\argmin}{arg\,min}
\DeclarePairedDelimiter{\norm}{\lVert}{\rVert}

\begin{document}
\title{Adaptive variational quantum eigensolvers for highly excited states}

\author{Feng Zhang}
\affiliation{Ames Laboratory, Ames, Iowa 50011, USA}

\author{Niladri Gomes}
\affiliation{Ames Laboratory, Ames, Iowa 50011, USA}

\author{Yongxin Yao}
\affiliation{Ames Laboratory, Ames, Iowa 50011, USA}
\affiliation{Department of Physics and Astronomy, Iowa State University, Ames, Iowa 50011, USA}

\author{Peter P.~Orth}
\email{porth@iastate.edu}
\affiliation{Ames Laboratory, Ames, Iowa 50011, USA}
\affiliation{Department of Physics and Astronomy, Iowa State University, Ames, Iowa 50011, USA}

\author{Thomas Iadecola}
\email{iadecola@iastate.edu}
\affiliation{Ames Laboratory, Ames, Iowa 50011, USA}
\affiliation{Department of Physics and Astronomy, Iowa State University, Ames, Iowa 50011, USA}

\begin{abstract}
Highly excited states of quantum many-body systems are central objects in the study of quantum dynamics and thermalization that challenge classical computational methods due to their volume-law entanglement content. In this work, we explore the potential of variational quantum algorithms to approximate such states. We propose an adaptive variational algorithm, adaptive VQE-X, that self-generates a variational ansatz for arbitrary eigenstates of a many-body Hamiltonian $H$ by attempting to minimize the energy variance with respect to $H$. We benchmark the method by applying it to an Ising spin chain with integrable and nonintegrable regimes, where we calculate various quantities of interest, including the total energy, magnetization density, and entanglement entropy. We also compare the performance of adaptive VQE-X to an adaptive variant of the folded-spectrum method. For both methods, we find a strong dependence of the algorithm's performance on the choice of operator pool used for the adaptive construction of the ansatz. In particular, an operator pool including long-range two-body gates accelerates the convergence of both algorithms in the nonintegrable regime. We also study the scaling of the number of variational parameters with system size, finding that an exponentially large number of parameters may be necessary to approximate individual highly excited states. Nevertheless, we argue that these methods lay a foundation for the use of quantum algorithms to study finite-energy-density properties of many-body systems.
\end{abstract}

\date{\today}

\maketitle

\section{Introduction}
\label{sec: Introduction}
Quantum simulation is one of the most promising use cases of near-term quantum computers, with applications ranging from quantum chemistry to many-body quantum dynamics. Despite substantial recent advances in quantum hardware design and control, present-day noisy intermediate-scale quantum (NISQ) devices can execute only modestly deep quantum circuits with high fidelity~\cite{preskillQuantumComputingNISQ2018,Bharti21NISQReview}. In order to achieve quantum simulation beyond classically tractable regimes with NISQ hardware, it is thus necessary to devise quantum algorithms that harness the power of quantum mechanics while maintaining a limited circuit depth.

Variational hybrid quantum-classical algorithms~\cite{cerezo2020variational} aim to achieve this by mapping a problem of interest to the optimization of a cost function over a set of variational parameters $\bm\theta=(\theta_1,\dots,\theta_{\mathcal{N}})$ in an ansatz state $\ket{\psi(\bm\theta)}$ that can be prepared by a finite-depth circuit. These algorithms employ feedback between quantum and classical hardware, using the former to evaluate the cost function and the latter to update the variational parameters for the next evaluation. A promising example of such an algorithm is the Variational Quantum Eigensolver (VQE)~\cite{peruzzoVariationalEigenvalueSolver2014}, which finds a variational approximation of the ground state of a quantum system by minimizing the energy $E(\bm\theta)=\bra{\psi(\bm\theta)}H\ket{\psi(\bm\theta)}$, where $H$ is the Hamiltonian of the system of interest. The VQE has been used in a variety of contexts from quantum chemistry~\cite{peruzzoVariationalEigenvalueSolver2014,McClean16,kandalaHardwareefficientVariationalQuantum2017,omalleyScalableQuantumSimulation2016,collessComputationMolecularSpectra2018,grimsleyAdaptiveVariationalAlgorithm2019} and many-body physics~\cite{weckerProgressPracticalQuantum2015,hoEfficientVariationalSimulation2019,Matos21} to lattice gauge theories~\cite{Klco18,kokail2019self}.

Most applications of VQE-like algorithms to date have focused on ground states and low-lying excited states, e.g. by preparing variational states that span the low-energy manifold~\cite{mccleanHybridQuantumclassicalHierarchy2017,stairMultireferenceQuantumKrylov2020,nakanishiSubspacesearchVariationalQuantum2019,parrishQuantumComputationElectronic2019}, or by adding penalty projector terms to the Hamiltonian cost function that project out lower energy states~\cite{higgottVariationalQuantumComputation2019}.  However, the ground and low-lying excited states generally have limited entanglement content and thus are more amenable to classical simulation than generic quantum states. For example, ground states of gapped local Hamiltonians in 1D are known to exhibit area law entanglement~\cite{hastings2007area}, a fact that underlies the success of the classical density matrix renormalization group (DMRG) algorithm in such systems~\cite{White92,schollwockDensitymatrixRenormalizationGroup2011}. In contrast, \textit{highly excited} eigenstates of such Hamiltonians, i.e. ones near the middle of the many-body spectrum, are generically believed to satisfy the eigenstate thermalization hypothesis (ETH)~\cite{Deutsch1991,Srednicki1994,Rigol2008,DAlessio2016,Deutsch18}. Highly excited states that obey the ETH exhibit extensive ``volume-law" entanglement~\cite{Deutsch10,Garrison18} approaching that of a random state~\cite{pageAverageEntropySubsystem1993}, which limits classical computational studies of such states to exact diagonalization (ED) of relatively small systems. Properties of (highly) excited states play an important role in finite-temperature and quantum-dynamical behavior. Probing excited states as a function of energy density has been used to study many-body mobility edges~\cite{Luitz2015,Naldesi16,Nag17,Wei19,Zhang21} and emulsions~\cite{Schecter2018} (also posing various fundamental open questions~\cite{DeRoeck2016,Crowley20,Brighi20}), as well as Hilbert-space fragmentation~\cite{Sala19,Khemani19} and rare low-entanglement ``many-body scar" states~\cite{Turner2018,Moudgalya18,Schecter2019}. Simulating highly excited states using quantum algorithms thus offers a promising venue for potential quantum advantage.

In this work, we study the capacity of VQE-like algorithms to find generic eigenstates of local quantum many-body Hamiltonians. A number of variational algorithms capable of searching for arbitrary eigenstates have been proposed. For example, the folded-spectrum method (FSM) applies the VQE algorithm to the nonlocal Hamiltonian $(H-\lambda)^2$ in order to find the eigenstate with energy closest to $\lambda$, which is a free parameter~\cite{macdonald1934modified,wang1994solving,peruzzoVariationalEigenvalueSolver2014,McClean16,WAVES}. Another approach is to use a cost function besides the energy, such as the energy variance $\braket{H}^2-\braket{H^2}$, which vanishes only for eigenstates of $H$~\cite{varianceVQE}. However, these algorithms have so far only been used to probe low-lying excited states. Moreover, they have typically been applied with fixed variational ans\"atze used for ground-state problems, which may or may not be able to provide faithful low-depth representations of highly excited states.

\begin{figure}[t!]
    \centering
    \includegraphics[width=\columnwidth]{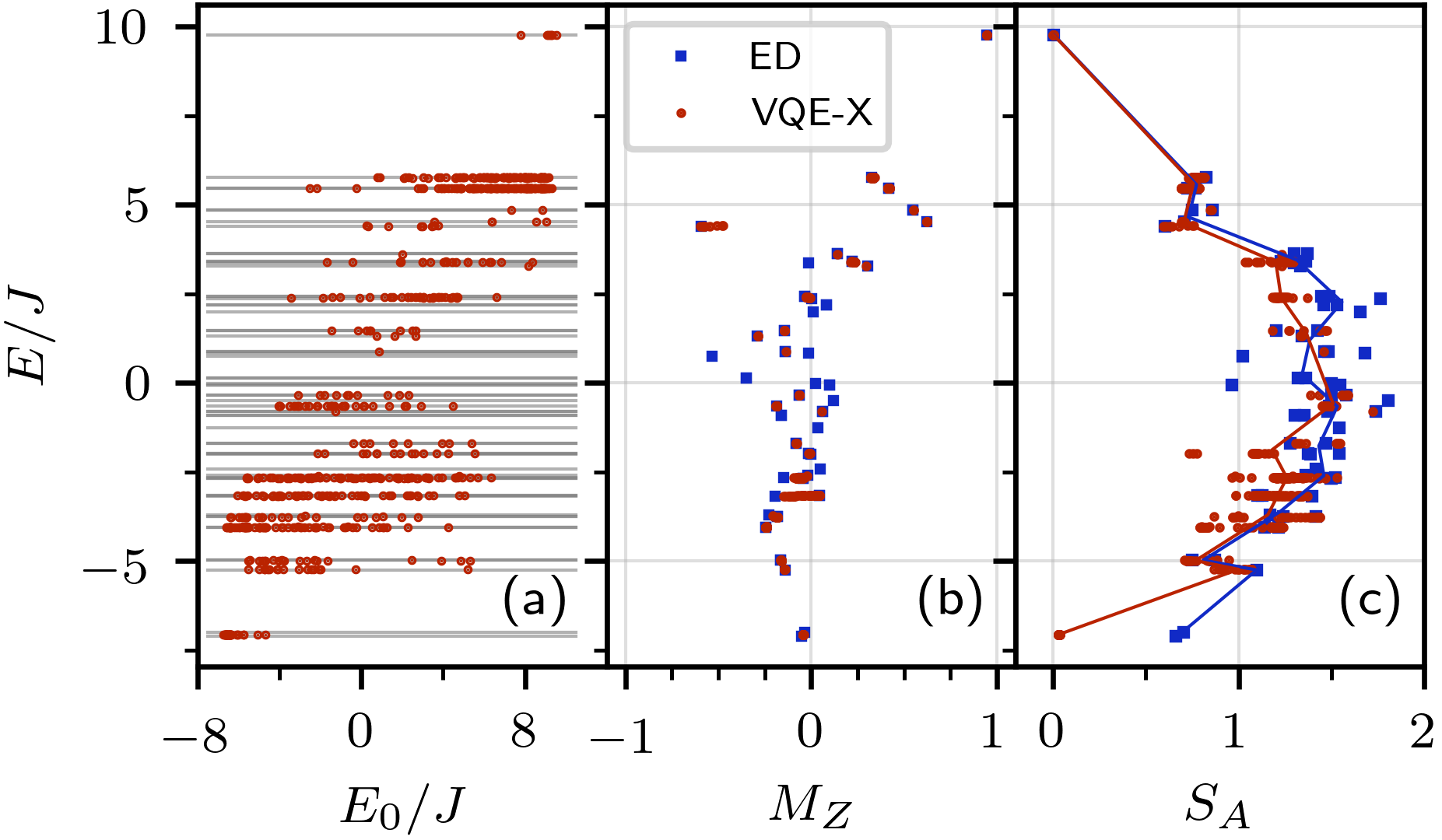}
    \caption{Application of adaptive VQE-X to the nonintegrable mixed-field Ising model (MFIM) at system size $N=6$, $h_x/J=0.8$, and $h_z/J=0.5$. (a) Scatter plot of the energy expectation value $E=\braket{\psi(\bm\theta)|H|\psi(\bm\theta)}$ of converged adaptive VQE-X trials against the energy expectation value $E_0$ of the initial product state $\ket{\psi_0}$ [Eq.~\eqref{eq:psi_0}]. Horizontal lines indicate the energies of exact eigenstates from ED. (b) Scatter plot of $E$ against the magnetization density $M_Z$ [Eq.~\eqref{eq:Mz}] for exact eigenstates from ED (blue points) and adaptive VQE-X trial eigenstates (red points). (c) Scatter plot of $E$ against the half-chain entanglement entropy $S_A$ [Eq.~\eqref{eq:SA}]. Solid red and blue lines connect energy-bin-averaged $S_A$ points as discussed below Eq.~\eqref{eq:SA}.}
    \label{fig:1}
\end{figure}

In Sec.~\ref{sec: VQE-X Algorithm}, we propose an algorithm, adaptive VQE-X, that samples eigenstates from the whole spectrum of $H$ in an unbiased manner using the energy variance as a cost function. Inspired by adaptive variational approaches to ground-state~\cite{grimsleyAdaptiveVariationalAlgorithm2019,qubit-ADAPT,claudinoBenchmarkingAdaptiveVariational2020,Ryabinkin2020,AVQITE} and time-evolution problems~\cite{AVQDS}, the algorithm avoids the pitfall of a fixed variational ansatz by adaptively constructing its own ansatz using operators from a predefined pool. The adaptive procedure is carried out alongside the variance optimization in such a way that new operators are added to the ansatz only if they accelerate the algorithm toward convergence. The algorithm is applied to an ensemble of random product states and the set of converged trials yields a sampling of approximate eigenstates from the whole spectrum, which we envision can then be used to study finite-energy-density properties of quantum many-body systems (see, e.g., Fig.~\ref{fig:1}).

In Sec.~\ref{sec: Application: Mixed-Field Ising Model}, we test the adaptive VQE-X algorithm on a benchmark quantum many-body system, the nonintegrable mixed-field Ising model (MFIM), of which the integrable transverse-field Ising model (TFIM) is a special case. We find that the algorithm succeeds in producing uniform coverage of the energy spectrum for small systems, and that the main performance bottleneck comes from the high-dimensional classical optimization of the variational parameters. Intriguingly, we find that the algorithm's success in probing eigenstates of integrable vs.~nonintegrable models depends on the choice of operator pool for the adaptive ansatz, with nonintegrable models benefiting substantially from the inclusion of longer-ranged two-body gates. We interpret this as a practical manifestation of the increased complexity of highly excited states in such systems relative to ground states and low-lying excitations above them.

In Sec.~\ref{sec: Comparison to Adaptive Folded-Spectrum Method}, we introduce a variant of the FSM that uses the adaptive procedure from VQE-X to approximate highly excited states near a fixed target energy. We show that this method results in variational circuits with a similar number of parameters to those generated within adaptive VQE-X. We close with a discussion in Sec.~\ref{sec: Discussion and Outlook}, including a scaling analysis of the complexity of adaptive VQE-X circuits, and use this to motivate directions for future work.

\section{Adaptive VQE-X Algorithm}
\label{sec: VQE-X Algorithm}
In order to prepare generic excited states $\ket{n}$ of a Hamiltonian $H$ for a system of $N$ qubits, we use the energy variance,
\begin{equation}
 \mathcal{C}(\ket{\psi(\boldsymbol{\theta}}) = \braket{\psi(\boldsymbol{\theta}) | H^2 | \psi(\boldsymbol{\theta})} - \braket{\psi(\boldsymbol{\theta}) | H | \psi(\boldsymbol{\theta})}^2 \,,
 \label{eq:cost_energy_variance}
\end{equation}
as a cost function.
Here, $\ket{\psi(\boldsymbol{\theta})} = U(\boldsymbol{\theta})\ket{\psi_0}$ is a variational state obtained by applying the unitary circuit $U(\boldsymbol{\theta})$ with parameters $\theta_\alpha \in [0, 2 \pi)$ to an initial state $\ket{\psi_0}$.
Importantly, this cost function vanishes for \emph{any} energy eigenstate. To construct the unitary $U(\boldsymbol{\theta})$ we utilize an adaptive approach that starts from a random initial product state 
\begin{align}
\label{eq:psi_0}
\ket{\psi_0} = \bigotimes_{i=1}^N ( \cos \varphi_i \ket{0}_i + \sin \varphi_i \ket{1}_i ),    
\end{align}
where the angles $\varphi_i$ are drawn uniformly from some distribution.
We then iteratively build the ansatz as follows: at step $\alpha \in [1, \mathcal{N}_{\text{max}}]$ we add an operator $e^{i \theta_\alpha \mathcal{O}_\alpha}$, where $\mathcal{O}_\alpha$ is drawn from a predefined pool $\mathscr{P}$ of Pauli string operators. We choose to add the operator that results in the minimal cost function, i.e.,
\begin{subequations}
\begin{equation}
 \mathcal{O}_\alpha = \argmin_{\{\mathcal{O}_\nu\}} \min_{\theta_\nu} \mathcal{C}\bigl (e^{i \theta_\nu \mathcal{O}_\nu} \ket{\psi_{\alpha - 1}} \bigr) \,,
    \label{eq:op_selection}
\end{equation}
where 
\begin{align}
    \ket{\psi_\alpha} =\ket{\psi(\bm\theta_\alpha)}= \prod_{\nu = 1}^\alpha e^{i \theta_\nu \mathcal{O}_\nu} \ket{\psi_0}
\end{align}
\end{subequations}
and we keep all parameters in $\ket{\psi_{\alpha -1}}$ fixed in Eq.~\eqref{eq:op_selection}. Since we avoid evaluating the cost-function gradient throughout the algorithm, the operator $\mathcal O_\alpha$ is selected based on one-parameter optimizations, which can be efficiently performed using a Golden-section search algorithm. We then update all parameter values $\boldsymbol{\theta}_\alpha = \{\theta_1, \ldots, \theta_\alpha\}$ by optimizing the enlarged ansatz: $\boldsymbol{\theta}_\alpha = \argmin_{\boldsymbol{\theta}_\alpha} \mathcal{C}(\ket{\psi_\alpha})$. We proceed to the next step $\alpha \rightarrow \alpha + 1$ unless we have reached the maximal number of steps $\alpha = \mathcal{N}_{\text{max}}$ or achieved the convergence criterion $\mathcal F(\ket{\psi_\alpha}) < \delta$, where
\begin{align}
    \mathcal F(\ket{\psi(\bm\theta)})=1 - \frac{|\braket{\psi(\bm\theta) | H | \psi(\bm\theta)}|}{\norm{H \ket{\psi(\bm\theta)}}}.
    \label{eq:convergence_criterion}
\end{align}
In practice, we first calculate $\norm{H \ket{\psi(\bm\theta)}}$ to avoid the possible divergence of $\mathcal F(\ket{\psi(\bm\theta)})$. The algorithm is terminated if $\norm{H \ket{\psi(\bm\theta)}} < \delta$, which indicates that $\ket{\psi(\bm\theta)}$ converges to an eigenstate with $E=0$. Otherwise, the convergence criterion in Eq.~\eqref{eq:convergence_criterion} evaluates to $0$ if and only if $H\ket{\psi(\bm\theta)}$ is collinear with $\ket{\psi(\bm\theta)}$. Note that $\norm{H \ket{\psi(\bm\theta)}}=\sqrt{\braket{\psi(\bm\theta)|H^2|\psi(\bm\theta)}}$, so $\mathcal F(\ket{\psi(\bm\theta)})$ can be computed using the same expectation values that yield the cost function $\mathcal C(\ket{\psi(\bm\theta)})$. Indeed, $\mathcal C$ and $\mathcal F$ are closely related and vanish identically on the same set of states. $\mathcal F$ is more natural as a convergence criterion because it is a dimensionless and intensive quantity (i.e., it does not scale with system size $N$), unlike the energy variance $\mathcal C$. We denote the step at which convergence is achieved by $\alpha=\mathcal N_{\rm c}$, which counts the number of variational parameters needed to achieve convergence.


The adaptive VQE-X algorithm assumes that the variance cost function $\mathcal C$ can be measured efficiently on a quantum processing unit (QPU). If $H$ is a sum of $O(N)$ Pauli strings, measuring $\mathcal C$ requires the evaluation of $O(N^2)$ Pauli-string expectation values. Each Pauli-string measurement can be converted to a computational basis (CB) measurement using at most $N$ one-qubit gates. Thus, compared to a VQE calculation using the same ansatz, which would require measuring $O(N)$ Pauli strings, cost-function evaluation in adaptive VQE-X requires only a polynomial quantum-resource overhead.

Optimization calculations, e.g. for the ansatz parameters $\bm\theta$ and for determining the choice of $\mathcal O_\alpha$ at each step, are carried out on classical hardware. Any optimization algorithm can be used, but we opt for the Nelder-Mead algorithm as it avoids the need to measure gradients of the cost function with respect to the ansatz parameters.

Before applying adaptive VQE-X to a specific model, we point out a potential pitfall of the convergence criterion $\mathcal F(\ket{\psi(\bm\theta)})<\delta$, with $\mathcal F$ defined in Eq.~\eqref{eq:convergence_criterion}. It is possible for a state $\ket{\Psi}$ that is a superposition of eigenstates nearby in energy to satisfy $\mathcal F(\ket{\Psi})<\delta$. For example, if we write $\ket{\Psi}=a\ket{E}+b\ket{E+\Delta}$, with $\ket{E}$ and $\ket{E+\Delta}$ energy eigenstates and $|a|^2+|b|^2=1$, and work to lowest order in the energy difference $\Delta$, we find that
\begin{align}
    \mathcal F(\ket{\Psi})\approx \frac{|a|^2|b|^2}{2E^2}\Delta^2.
    \label{eq:5}
\end{align}
Since the typical energy spacing between consecutive eigenstates in the bulk of the many-body energy spectrum is $\Delta\sim e^{-cN}$ for some constant $c>0$, this means that it is possible for a converged trial to yield a superposition of nearby energy eigenstates. We will see consequences of this fact in Sec.~\ref{sec: Application: Mixed-Field Ising Model}. Given the exponential density of states at finite energy density, this drawback is likely encountered by any cost function capable of identifying individual eigenstates (see, e.g., Sec.~\ref{sec: Discussion and Outlook}). For certain applications, however---e.g., the use of converged adaptive VQE-X trials to estimate microcanonical averages, as discussed in Sec.~\ref{sec: Discussion and Outlook}---the fact that the algorithm can converge to superpositions may not be detrimental.

\section{Application: Mixed-Field Ising Model}
\label{sec: Application: Mixed-Field Ising Model}
We apply the adaptive VQE-X algorithm to finding highly excited states of the mixed-field Ising model (MFIM),
\begin{align}
\label{eq:H}
   H=J\sum^N_{i=1} Z_iZ_{i+1}+\sum^N_{i=1} \left( h_x\, X_i+h_z\, Z_i\right),
\end{align}
which for generic parameter values is a paradigmatic example of a Hamiltonian believed to satisfy the ETH~\cite{Kim13,Kim14}. A special case of the model is the integrable transverse-field Ising model (TFIM), which is often used as a benchmark for quantum algorithms. We work with periodic boundary conditions (PBC) such that $N+1\equiv 1$. We focus in this section on the $N=6$ case, and discuss the scaling with $N$ of the algorithm's performance in Sec.~\ref{sec: Discussion and Outlook}.

Since excited states of integrable models are believed to be less complex than those of nonintegrable models, we will be interested in comparing the performance of the adaptive VQE-X algorithm in integrable and nonintegrable regimes. In units where $J=1$, parameter values that we take to represent the integrable and nonintegrable cases are $(h_x,h_z)=(0.8,0)$ and $(0.8,0.5)$, respectively. We will also examine the dependence of the algorithm's performance on the choice of operator pool $\mathscr{P}$ from which the operators $\mathcal O_\alpha$ are drawn at each step. We make use of two operator pools: the ``minimal" pool~\cite{qubit-ADAPT}
\begin{subequations}
\begin{align}
    \mathscr{P}_{\rm min}=\{Y_i\}^N_{i=1}\cup\{Y_{i}Z_{i+1}\}^N_{i=1}
\end{align}
and the ``maximal" pool
\begin{align}
    \mathscr{P}_{\rm max}=\{Y_i\}^N_{i=1}\cup\{Y_{i}Z_{j}\}^N_{i,j=1}\cup\{Y_{i}X_{j}\}^N_{i,j=1},
\end{align}
\end{subequations}
which, like the Hamiltonian~\eqref{eq:H}, are defined with PBC. For a given system size $N$, we can divide our numerical tests into four categories defined by whether integrable/nonintegrable parameters and pools $\mathscr{P}_{\rm min}$/$\mathscr{P}_{\rm max}$ are used.

For each choice of $N,(h_x,h_z),$ and $\mathscr P$, our implementation of the algorithm consists of many ``trials'', each starting from a different initial product state $\ket{\psi_0}$. The distribution of initial product states is not strictly uniform---rather, we generate product states using uniform distributions for the angles $\varphi_i$ and then postselect to obtain an ensemble $\{\ket{\psi_0}\}$ with a roughly uniformly distributed average energy $\{\braket{\psi_0|H|\psi_0}$\}. As the average energy of the initial and final states are (weakly) correlated, see Fig.~\ref{fig:1}(a), this postselection helps to ensure a more uniform coverage of the energy spectrum. We used an ensemble of $\sim1000$ initial product states and kept only the trials that converged in at most $\mathcal{N}_{\text{max}}=100$ steps.

\begin{figure}[t]
    \centering
    \includegraphics[width=\linewidth]{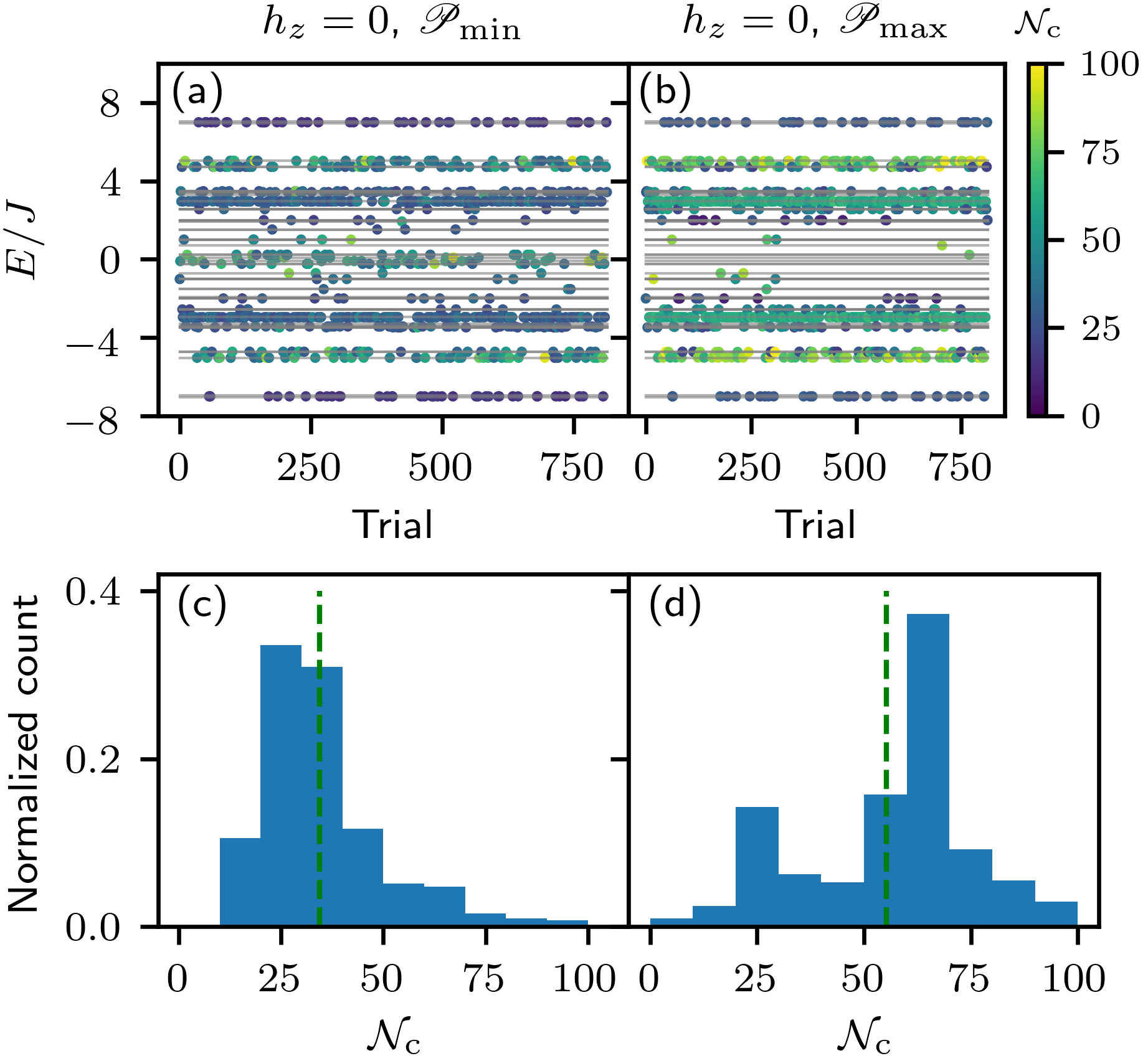}
    \caption{Adaptive VQE-X results for the integrable TFIM at system size $N=6$, $h_x/J=0.8$ and $h_z = 0$. (a, b) Energy of the converged variational state $\ket{\psi_{\mathcal{N}_c}}_n$ versus trial number $n$ labeling different random initial states $\ket{\psi_0(\boldsymbol{\varphi})}_n$. Data points are colored according to the length $\mathcal N_{\rm c}(n)$ of the converged adaptive ansatz. Horizontal lines indicate the energies of exact eigenstates from ED. Panel (a) is for the minimal pool, and panel (b) is for the maximal pool as described in the text. We observe that convergence is more uniform over the spectrum for the minimal pool. 
    On average, for converged trials using the minimal pool, we find that $\mathcal N_{c}$ is $42\pm 12$ for excited states, compared to $15\pm 1.5$ for the ground and highest excited states.
    (c, d) Histogram of circuit lengths of the converged adaptive ans\"atze, normalized to the total number of converged trials. Panel (c) [(d)] is for minimal [complete] pool. Green dashed line denotes arithmetic average over all converged trials, which shows that the minimal pool produces shorter ans\"atze by a factor of $1.6$. }
    \label{fig:results_integrable}
\end{figure}

Our results for the integrable TFIM are shown in Fig.~\ref{fig:results_integrable}. Each point in Fig.~\ref{fig:results_integrable}(a),(b) represents a trial that converged to an eigenstate to within a tolerance $\delta=10^{-4}$ (see Eq.~\eqref{eq:convergence_criterion}). The $x$ coordinate of each point represents the trial number, with each trial corresponding to a different random initial product state. The $y$ coordinate represents the final energy of the converged state. The color of each point denotes the number $\mathcal N_{\rm c}$ of variational parameters for that trial, with lighter colors denoting more parameters. The minimal pool $\mathscr{P}_{\rm min}$ produces more even coverage of the spectrum in this case. Trials making use of the maximal pool $\mathscr P_{\rm max}$ are much less likely to converge to states in the middle of the spectrum. Moreover, trials using $\mathscr P_{\rm max}$ require more variational parameters. The distribution of $\mathcal N_{\rm c}$ over trials for the two pools is shown in Fig.~\ref{fig:results_integrable}(c),(d). For $N=6$, the average $\mathcal N_{\rm c}$ value for $\mathscr P_{\rm min}$ trials is roughly half that of the $\mathscr P_{\rm max}$ trials. 

We attribute the success of the $\mathscr{P}_{\rm min}$ trials in the integrable case to the exact solvability of the TFIM. Since the TFIM maps to a model of free fermions, all eigenstates are product states in the momentum-space occupation number basis. Intuitively, representing such states should not require long-range two-body Hermitian generators of the type present in $\mathcal P_{\rm max}$. In this case, including such operators in the operator pool simply generates more possible directions in which to expand the ansatz, complicating the optimization procedure and hindering convergence.

\begin{figure}[t]
    \centering
    \includegraphics[width=\linewidth]{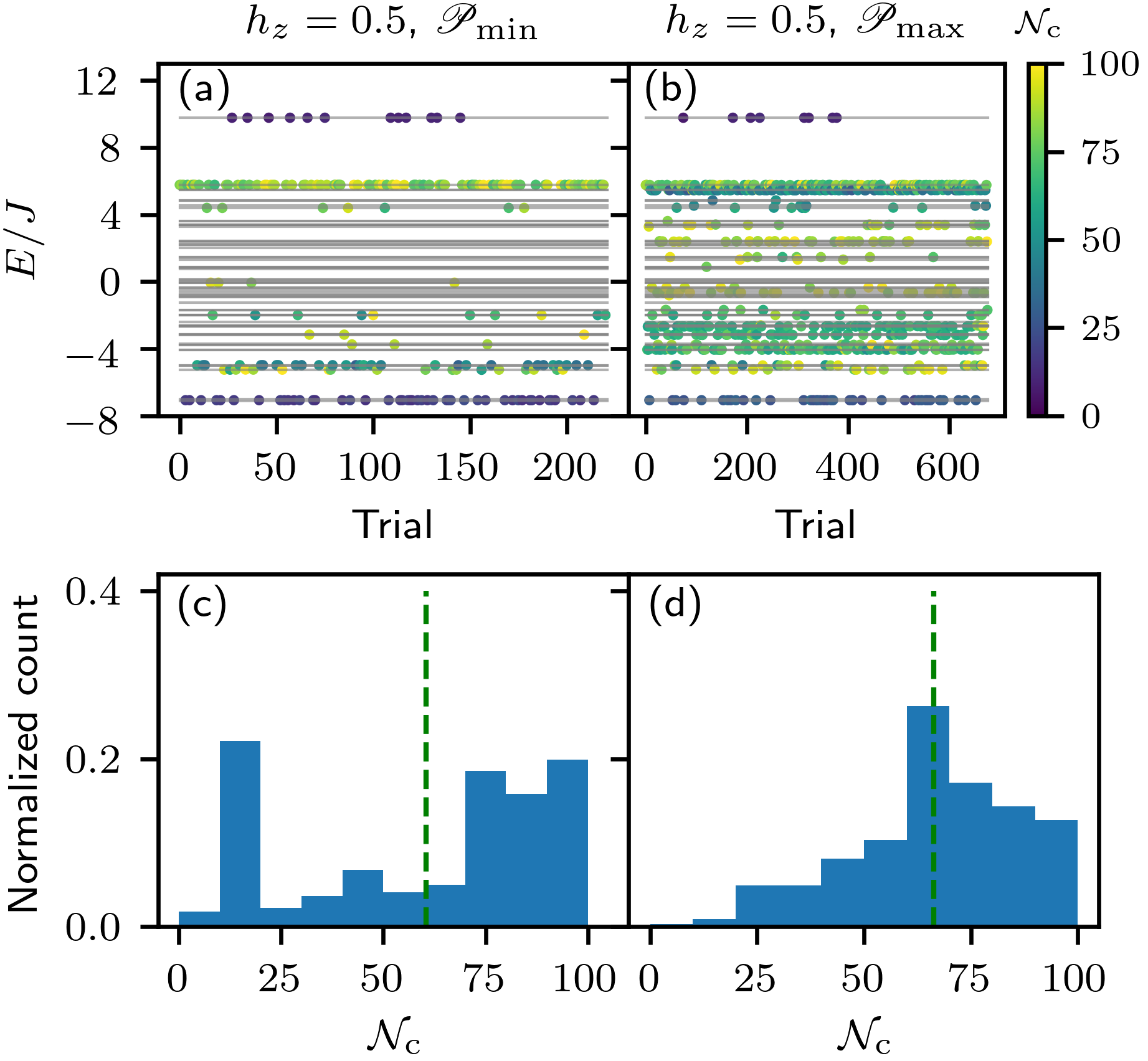}
    \caption{Adaptive VQE-X results for the nonintegrable MFIM at $N=6$, $h_x/J=0.8$, and $h_z/J=0.5$. (a)--(d) are analogous to the corresponding panels of Fig.~\ref{fig:results_integrable}. Comparing (a) and (b), we see that the pool $\mathscr{P}_{\rm max}$ now outperforms $\mathscr{P}_{\rm min}$ in producing more even coverage of the energy spectrum. In (d), we see that the distribution of $\mathcal N_{\rm c}$, the number of variational parameters, for the adaptive circuits drawn from $\mathscr P_{\rm max}$, is unimodal with an arithmetic average over all converged trials $\sim65$ (green dashed line). In contrast, in (c), we see that the same distribution for $\mathscr P_{\rm min}$ is bimodal, indicating that the minimal pool is unable to reach certain excited states accessible with $\mathscr P_{\rm max}$ with the same number of variational parameters. 
    On average, for converged trials using the maximal pool, we find that $\mathcal N_{c}$ is $75\pm 13$ for excited states, compared to $18\pm 2$ for the ground and highest excited states.}
    \label{fig:results_non_integrable}
\end{figure}

Intriguingly, we find that the scenario is reversed in the nonintegrable case, the results for which are shown in Fig.~\ref{fig:results_non_integrable}. In Fig.~\ref{fig:results_non_integrable}(a),(b) we see that the maximal pool $\mathscr{P}_{\rm max}$ generates both more converged trials and more even coverage of the full many-body spectrum than the minimal pool $\mathscr P_{\rm min}$. Evidently the long-range two-body generators present in $\mathscr{P}_{\rm max}$ facilitate faster convergence of each trial. This is consistent with the fact that the eigenstates of nonintegrable Hamiltonians generally exhibit volume-law entanglement; our results suggest that such complexity is more easily realized in the variational context by longer-range two-body gates.

While the coverage of exact eigenstates produced by adaptive VQE-X is generally quite uniform (if one uses the better performing pool), there are a few states that our algorithm did not converge to within the constraints of $\delta = 10^{-4}$ and maximal operator number $\mathcal{N}_{\text{max}}=100$ [see also Fig.~\ref{fig:1}(a)]. It is expected that by increasing the number of trials and by increasing $\mathcal{N}_{\text{max}}$ (which makes the classical optimization more time intensive), one can obtain full coverage. Another possibility is to soften the convergence criterion by increasing $\delta$, which, however, can result in a larger overlap of the converged trial wavefunctions with nearby energy eigenstates [see Eq.~\eqref{eq:5}].

The adaptive VQE-X algorithm produces variational states whose energy expectation values are close to those of exact eigenstates of $H$. It is natural to ask whether these variational states also capture other properties of the energy eigenstates, e.g. the expectation values of observables besides the Hamiltonian. To this end, we plot in Fig.~\ref{fig:1}(b) the expectation value in each converged variational state of the magnetization density
\begin{align}
\label{eq:Mz}
M_Z=\frac{1}{N}\sum^N_{i=1} Z_i.
\end{align}
Adaptive VQE-X data for the $N=6$ nonintegrable case using the pool $\mathscr P_{\rm max}$ are compared against ED. We find that the variational states provide excellent agreement with the exact results except near $E\approx -3$ and $E\approx5$, where VQE-X states for different trials provide a near continuum of magnetization densities. We have checked explicitly that the algorithm has converged to superpositions of a few nearby energy eigenstates in these cases; the dependence of $\braket{M_Z}$ on the superposition amplitudes yields the continuum of observed values near these energies.
\begin{figure}[tbh]
    \centering
    \includegraphics[width=\linewidth]{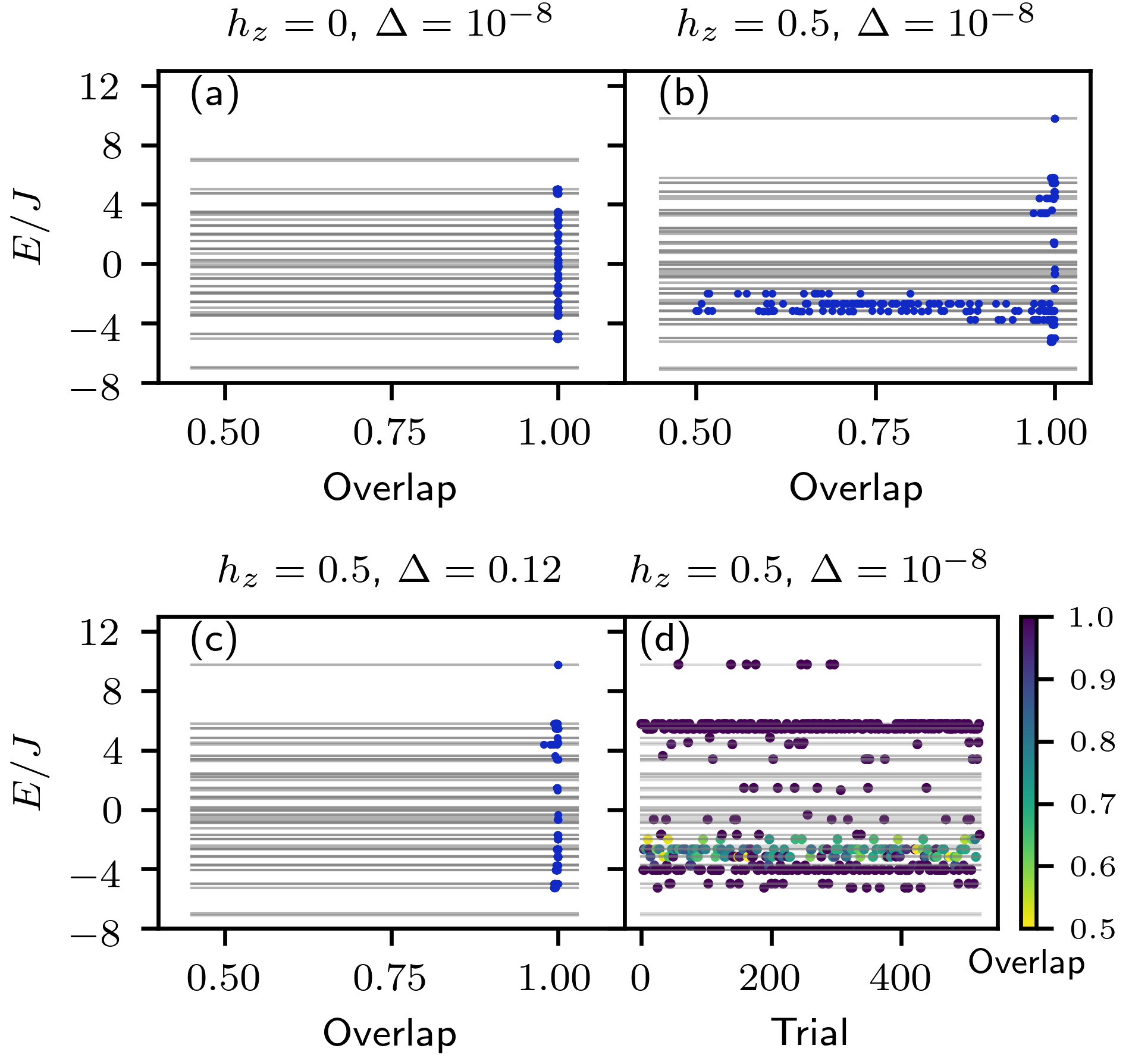}
    \caption{Overlap of adaptive VQE-X states with exact eigenstates in $N=6$ model for convergence threshold $\delta = 10^{-5}$. Panel (a) is for integrable model and panels (b-d) are for nonintegrable model with $h_z/J = 0.5$. We consider exact states with energy difference less than $\Delta/J$ as being degenerate and show the overlap onto such defined "degenerate" subspaces. (a) Overlap close to one shows that VQE-X produces proper eigenstates in the integrable model. (b-d) While most converged VQE-X states have overlap with a single energy eigenstate, some VQE-X states are superpositions of nearby energy eigenstates. In panel (c), we choose $\Delta/J = 0.12$, which corresponds to the small energy difference of the two cat-like ground states, to demonstrate that VQE-X states are composed of close-by energy eigenstates [in agreement with Eq.~\eqref{eq:5}]. To emphasize that the majority of trials converge to states having high overlap with a small number of eigenstates, panel (d) shows the same data as panel (b) resolved by trial number. Data points are color coded by their overlap with exact eigenstates in an energy window of width $\Delta$. }
    \label{fig:overlap}
\end{figure}

Another probe of the variational excited states produced by adaptive VQE-X is provided by the von Neumann entanglement entropy 
\begin{align}
\label{eq:SA}
S_A=-\text{tr}(\rho_A\ln\rho_A).
\end{align}
$S_A$ is defined with respect to a bipartition of the system into disjoint subsystems $A$ and $B$, such that the reduced density matrix $\rho_A=\text{tr}_B\rho$. In Fig.~\ref{fig:1}(c), we calculate $S_A$ for both the ED and VQE-X states in the $N=6$ nonintegrable case with pool $\mathscr P_{\rm max}$, taking region $A$ to consist of the first three sites of the chain. The agreement between the exact and variational results in this case is much poorer than for the magnetization density $M_Z$. This indicates that the majority of our adaptive VQE-X trials actually converge to superpositions of eigenstates that are nearby in energy. This is most prominent for the quasi-degenerate ground states, which resemble the antiferromagnetic cat states $|\pm\rangle = \frac{1}{\sqrt{2}} (|\uparrow \downarrow \uparrow \ldots\rangle \pm |\downarrow \uparrow \downarrow \ldots\rangle)$ and therefore have $S_A \approx \ln 2$. In contrast, the VQE-X algorithm converges to product states formed by the superposition of these two cat-like eigenstates, for which $S_A$ vanishes. 
This discrepancy could be remedied by restricting the variational ansatz to have a definite spatial reflection eigenvalue.
Nevertheless, the results obtained upon averaging the values of $S_A$ found within a fixed energy window for both the adaptive VQE-X and ED states agree relatively well. This can be seen in Fig.~\ref{fig:1}(c), where the solid red and blue lines connect points obtained by dividing the energy axis into bins of fixed width and computing the average of $E$ and $S_A$ over each bin for the adaptive VQE-X and ED states, respectively. These bin-averaged entanglement-vs.-energy curves agree reasonably well in the bulk of the spectrum, indicating that the ensemble of adaptive VQE-X states can reproduce average properties of exact eigenstates within an energy window.

To further investigate the nature of the converged adaptive VQE-X states, we compute their overlap with the exact eigenstates. As shown in Fig.~\ref{fig:overlap}, the converged VQE-X circuits produce proper eigenstates over the full spectrum in the integrable case. For the nonintegrable model at $h_z/J =0.5$, we find that some converged circuits produce states that are superpositions of a few nearby energy eigenstates. By projecting the variational states onto groups of nearby eigenstates within an energy window $\Delta$, we demonstrate that almost all the weight comes from states within $\Delta/J < 0.12$, which corresponds to the small energy difference between the two cat-like ground states of the model. We note that in the nonintegrable model the exact degeneracies that are present in the integrable case are split into nearby non-degenerate levels. The convergence to nearby states is expected from Eq.~\eqref{eq:5} and leads to the discrepancies between exact and VQE-X results in the magnetization and entanglement entropy reported in Fig.~\ref{fig:1}.

\section{Comparison to Adaptive Folded-Spectrum Method}
\label{sec: Comparison to Adaptive Folded-Spectrum Method}

The adaptive VQE-X algorithm formulated in Sec.~\ref{sec: VQE-X Algorithm} aims to provide uniform coverage of the energy spectrum by employing a random ensemble of initial states with approximately uniformly distributed energy expectation values. In practice, however, one might desire a variant of this algorithm that targets highly excited states near an arbitrary target energy $\lambda$. To this end, we introduce a variant of the FSM that uses the adaptive procedure outlined in Sec.~\ref{sec: VQE-X Algorithm} to self-generate a variational ansatz for excited states with energy near $\lambda$. The adaptive FSM is obtained from adaptive VQE-X by replacing the cost function in Eq.~\eqref{eq:cost_energy_variance} by 
\begin{equation}
    \mathcal{C}(\ket{\psi(\boldsymbol{\theta}}) = \braket{\psi(\boldsymbol{\theta}) | (H-\lambda)^2 | \psi(\boldsymbol{\theta})},
 \label{eq:cost_folded_spectrum}
\end{equation}
and leaving the remainder of the algorithm unchanged.
The new cost function drives the variational method to converge
to the eigenstate with energy closest to the shift parameter $\lambda$. 
Note that the quantum resource requirements for evaluating the FSM cost function \eqref{eq:cost_folded_spectrum} and the VQE-X cost function \eqref{eq:cost_energy_variance} on a QPU are comparable, since both cost functions require the evaluation of $\braket{\psi(\bm\theta)|H^2|\psi(\bm\theta)}$ and $\braket{\psi(\bm\theta)|H|\psi(\bm\theta)}$.

Eigenvalues and eigenvectors from throughout the many-body spectrum are computed by scanning through a range of $\lambda$ values and running this adaptive variational algorithm. To explore the full many-body spectrum, the range of $\lambda$ can be chosen to coincide with the bandwidth of $H$. In regimes where classical simulations are tractable, the bandwidth can be obtained using, e.g., the Lanczos method to obtain the lowest and highest eigenvalues of $H$. Alternatively, the bandwidth of $H$ can be estimated by performing a ``qubit mean-field" calculation \cite{Ryabinkin2018a,Ryabinkin2018b,Ryabinkin2020} wherein a set of variational parameters $\bm\varphi=(\varphi_1,\dots,\varphi_n)$ parameterizing a product state $\ket{\phi(\bm\varphi)}$ are optimized to estimate the minimal and maximal values of the cost function $\braket{\phi(\bm\varphi)|H|\phi(\bm\varphi)}$. 

\begin{figure}[t!]
    \centering
    \includegraphics[width=\linewidth]{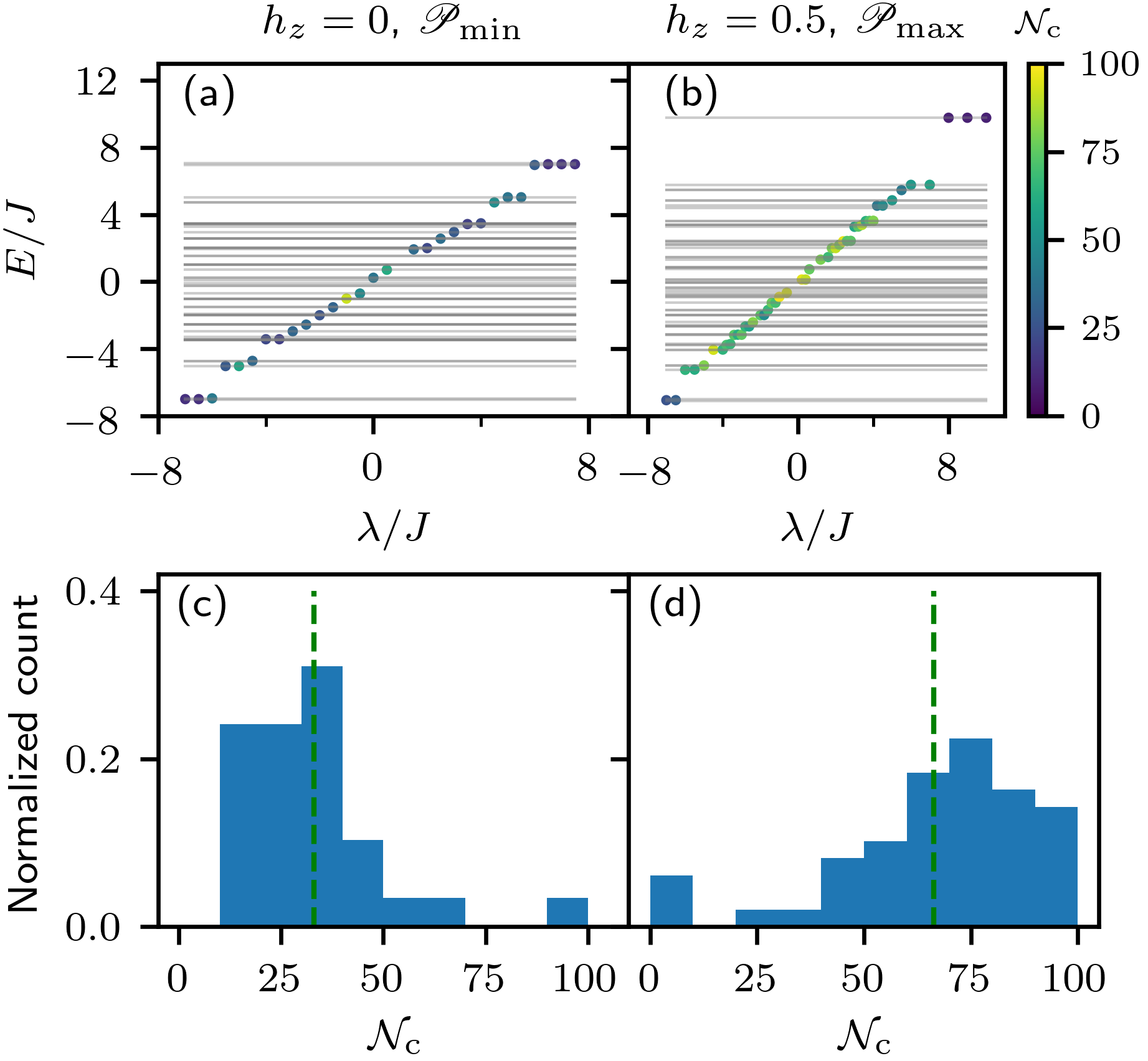}
    \caption{Adaptive FSM results for the MFIM with $N=6$ and $h_x/J=0.8$ in integrable (a,c) and nonintegrable (b,d) regimes. (a, b) Energy of the converged variational state $\ket{\psi_{\mathcal{N}_c}}_\lambda$ versus shift parameter $\lambda$, with data points color-coded by $\mathcal N_c$. Results for each $\lambda$ start from the same random initial product state. Panel (a) is for the minimal pool and $h_z=0$ and (b) is for the maximal pool and $h_z/J=0.5$. (c, d) Histogram of circuit lengths of the converged adaptive ans\"atze, normalized to the total number of converged runs. Panel (c) [(d)] is for minimal ($h_z/J=0.0$) [maximal ($h_z/J=0.5$)] pool. Green dashed lines denote the arithmetic average, which shows that the average number of variational parameters is similar to that of adaptive VQE-X.}
    \label{fig:results_fs}
\end{figure}

We have tested the adaptive FSM for both the integrable TFIM ($h_{z} = 0$) and the nonintegrable MFIM ($h_{z} = 0.5$) using the pools $\mathscr{P}_{\rm min}$ and $\mathscr{P}_{\rm max}$ for $N\leq 7$. Our results for the integrable and non-integrable regimes at $N=6$ are shown in Fig.~\ref{fig:results_fs}. We focus on results obtained using the pools $\mathscr{P}_{\rm min}$ for the integrable case and $\mathscr{P}_{\rm max}$ for the nonintegrable case, as these cases were found to provide the most favorable results for adaptive VQE-X in Sec.~\ref{sec: Application: Mixed-Field Ising Model}. Fig.~\ref{fig:results_fs}(a),(b) shows the dependence of the converged energy on $\lambda$. Each point in Fig.~\ref{fig:results_fs}(a),(b) represents a converged solution to an eigenstate to within a tolerance $\delta=10^{-4}$ (see Eq.~\eqref{eq:convergence_criterion}); we choose `one' of the same random initial product states that we used in the adaptive VQE-X method. We also tested the algorithm using qubit mean-field initial states, where we first minimize the FSM cost function over all possible $N$-qubit product states before running the adaptive FSM algorithm. We found the performance relative to a random initial state to be equivalent for the nonintegrable model and slightly worse for the integrable case. 
The color of each point denotes the number $\mathcal N_{\rm c}$ of variational parameters for the converged trial at that $\lambda$.  
When $\lambda$ is scanned over the full bandwidth of $H$, the pools $\mathscr{P}_{\rm min}$ and $\mathscr{P}_{\rm max}$ produce near-complete coverage of the spectrum within a limited number of variational steps, $\mathcal{N}_{\text{max}} =100$, for the integrable and nonintegrable cases, respectively. 

The results shown in Fig.~\ref{fig:results_fs}(a),(b) demonstrate that the adaptive FSM is capable of producing converged trials throughout the many-body spectrum for $N=6$ using the same operator pools and algorithmic parameters as adaptive VQE-X.
To compare the resulting variational circuits to the ones produced by adaptive VQE-X, we plot in Fig.~\ref{fig:results_fs}(c),(d) the distribution of the converged number of variational parameters, $\mathcal N_{\rm c}$, for the integrable and nonintegrable cases, respectively. The average value of $\mathcal N_{\rm c}$ (shown as green dashed lines) for each case are consistent with our adaptive VQE-X results for the same parameters and operator pools, see Figs.~\ref{fig:results_integrable} and \ref{fig:results_non_integrable}.
Moreover, we also observe the pool dependence noted in our discussion in Sec.~\ref{sec: Application: Mixed-Field Ising Model} of the adaptive VQE-X results for the same models. For example, we find that the minimal pool $\mathscr{P}_{\rm min}$ yields longer circuits and a reduced rate of convergence in the nonintegrable case, resulting in less uniform coverage of the spectrum (data not shown in Fig.~\ref{fig:results_fs}).

\section{Discussion and Outlook}
\label{sec: Discussion and Outlook}
\begin{figure}[t!]
    \centering
    \includegraphics{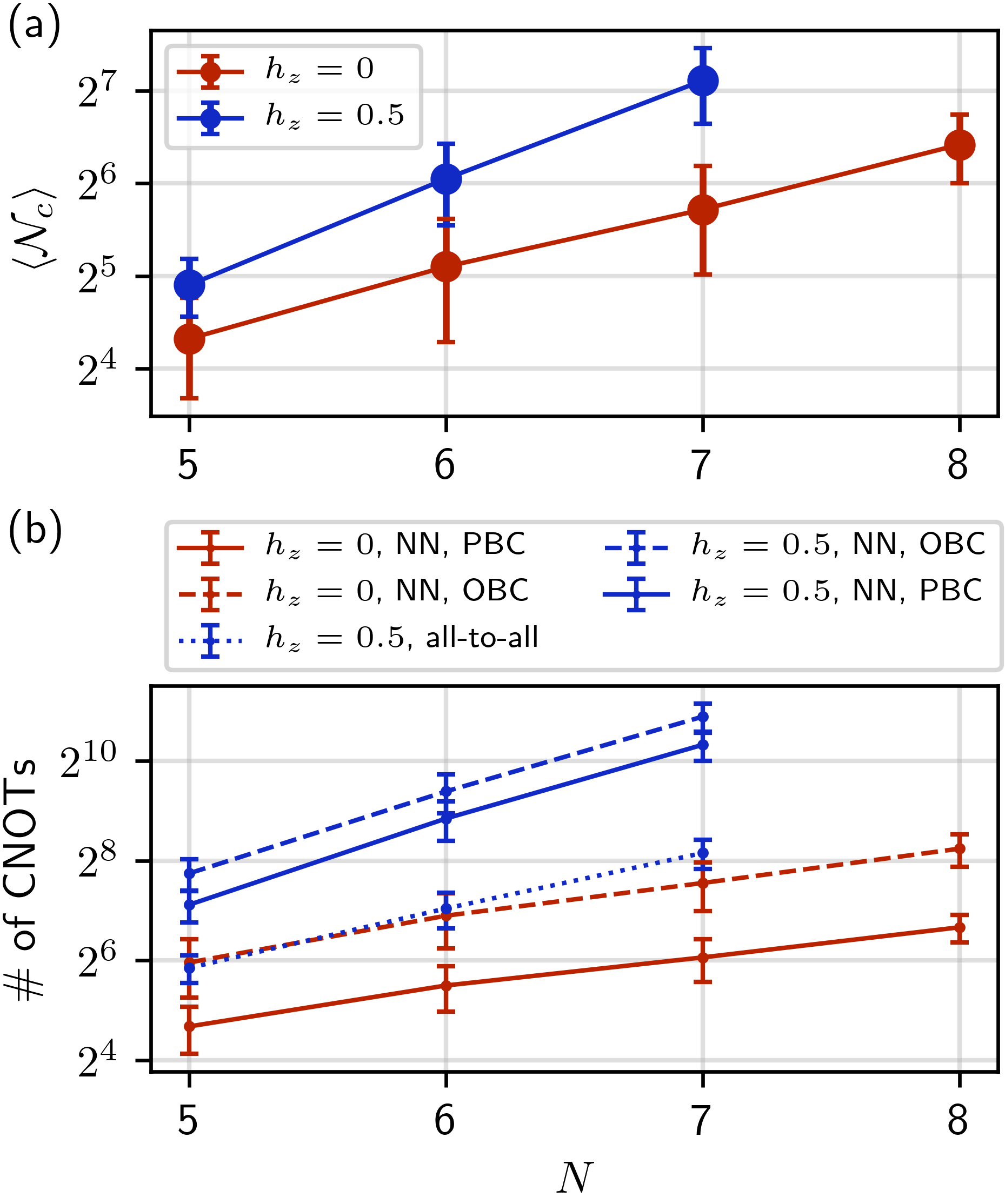}
    \caption{(a) Scaling of the average number of variational parameters $\langle \mathcal{N}_c \rangle$ (on a logarithmic scale) with system size $N$ for the integrable TFIM ($h_z = 0$) and nonintegrable MFIM ($h_z = 0.5$). Error bars indicate $\pm 1$ standard deviation around the mean of the distribution (the apparent asymmetry of the error bars is due to the logarithmic scale). We find that $\langle \mathcal{N}_c \rangle$ is larger and increases more rapidly for the nonintegrable model, where the scaling is approximately exponential and $\langle \mathcal{N}_c \rangle$ is close to the Hilbert space dimension $2^N$. 
    (b) Average number of CNOT gates in the converged adaptive VQE-X circuits for excited states as a function of system size $N$. Red (blue) lines denote the (non)integrable case with $h_z = 0$ ($h_z = 0.5$) and minimal (maximal) pool. Since the CNOT gate count depends on the connectivity of the qubits in the (hypothetical) QPU, we consider three cases: nearest-neighbor connectivity (NN) with open (OBC) or periodic (PBC) boundary conditions and all-to-all connectivity, as indicated in the figure. }
    \label{fig:scaling}
\end{figure}
Our results indicate that the adaptive VQE-X algorithm proposed in this work can be used to approximate highly-excited eigenstates of both integrable and nonintegrable Hamiltonians. However, the ans\"atze generated by the adaptive procedure contain many variational parameters. In the nonintegrable case at $N=6$ ($N=7$), the average number of variational parameters for a converged trial is $\braket{\mathcal N_c}\sim65$ ($\braket{\mathcal N_c}\sim 140$), which is comparable to the Hilbert-space dimension.
To investigate the system-size dependence of $\braket{\mathcal N_c}$, we performed adaptive VQE-X simulations for system sizes $N=5$--$8$ for both the integrable and nonintegrable cases. In Fig.~\ref{fig:scaling}(a), we plot the scaling with $N$ of $\braket{\mathcal N_c}$, which is the average of $\mathcal N_{c}$ over all converged trials at fixed $N$, on a semilog scale. We present results using the minimal (maximal) pool for the (non)integrable model. We find a clear trend towards exponential scaling for both the integrable and nonintegrable cases. Note that we only include data for $N=5$--$7$ for the nonintegrable case, due to an insufficient number of converged trials at $N=8$. 

In Fig.~\ref{fig:scaling}(b), we show the scaling with $N$ of the average number of CNOT gates ($N_{\text{CNOT}}$) in the converged VQE-X circuits. Different curves represent different (hypothetical) QPU connectivities as described in the caption. The average is performed over all trials that converged to excited states. It follows from panel (a) that $N_{\text{CNOT}}$ also increases exponentially with system size $N$. For the integrable model at $N=7$ and $h_z=0$, we find that the average number of CNOT gates is about 65 for nearest-neighbor (NN) connectivity with periodic boundary conditions (PBC) and about 200 for NN connectivity with open boundary conditions (OBC). Since the minimal pool only contains products of NN spin operators, the results are identical for NN PBC and all-to-all connectivity. In contrast, for the nonintegrable model at $N=7$ and $h_z/J=0.5$, where we employ the maximal pool, the average number of CNOT gates is substantially lower for all-to-all connectivity compared to NN PBC (NN OBC), specifically 280 versus 1300 (1900). This result suggests that a QPU platform that natively implements longer-range entangling gates (such as ion trap quantum computers) provides an advantage. 

Finally, we note that for comparison adaptive circuits for ground state preparation contain notably fewer CNOT operations, in particular at even $N$, where frustration due to the antiferromagnetic nature of the ground state is absent. 
For example, at $N=8, h_z=0$ ($N=6, h_z=0.5$) and PBC, we find the circuits to contain only $N_{\text{CNOT}} = 20$ $(120)$ CNOT gates for the ground state, in contrast to $N_{\text{CNOT}} = 100$ $(460)$ for excited states. This reflects the complexity of the wavefunction, which has only area-law entanglement for ground states, but follows a volume law for excited states in the non-integrable model. 

\begin{figure}[t!]
    \centering
    \includegraphics[width=\linewidth]{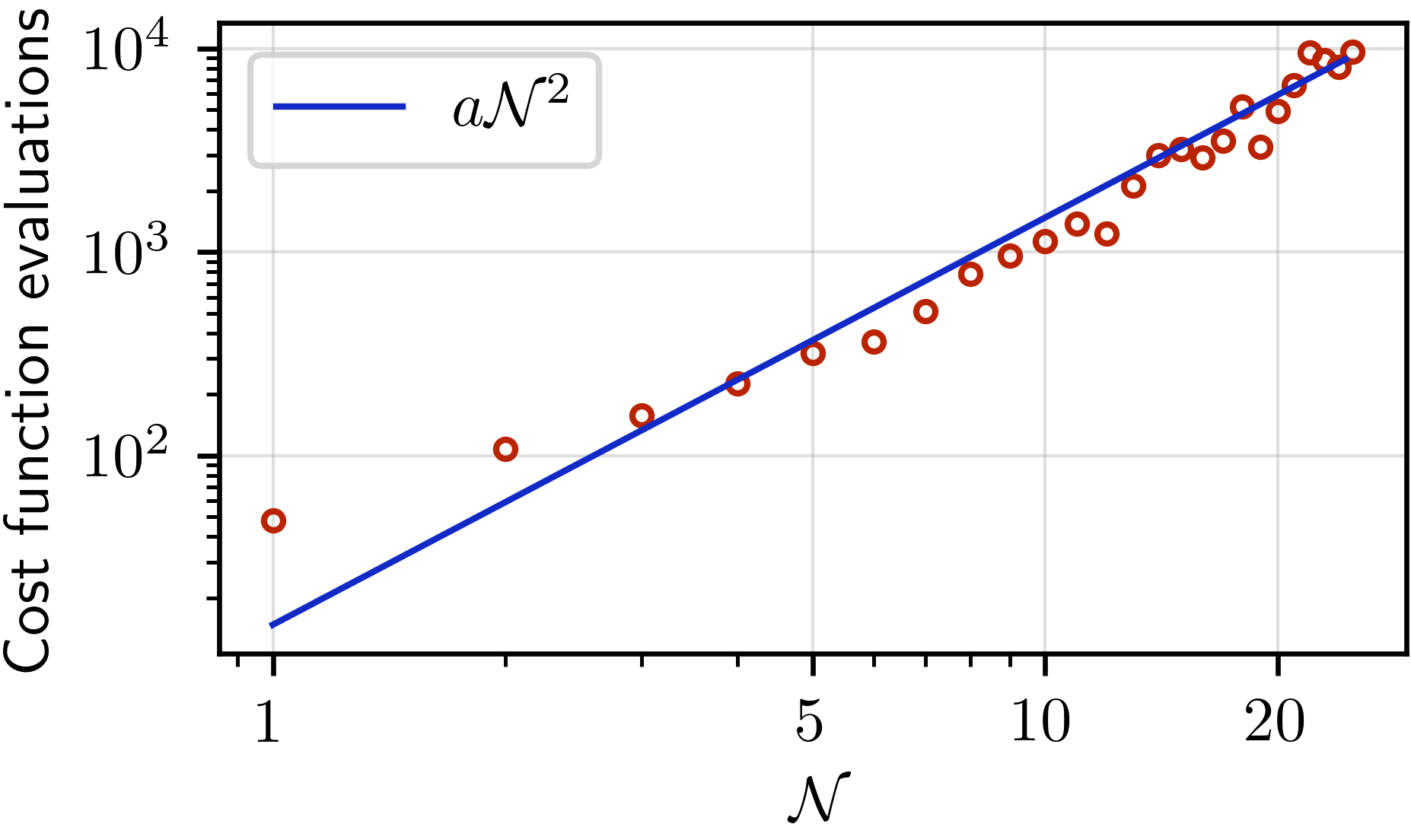}
    \caption{The number of cost function evaluations as a function of the number of variational parameters $\mathcal{N}$ for the early iterations of an application of adaptive VQE-X to the nonintegrable MFIM with $N=8$. The blue line is a quadratic fit to the data with $a = 14.8$.}
    \label{fig:func_eval}
\end{figure}

\begin{figure}[t!]
    \centering
    \includegraphics[width=\linewidth]{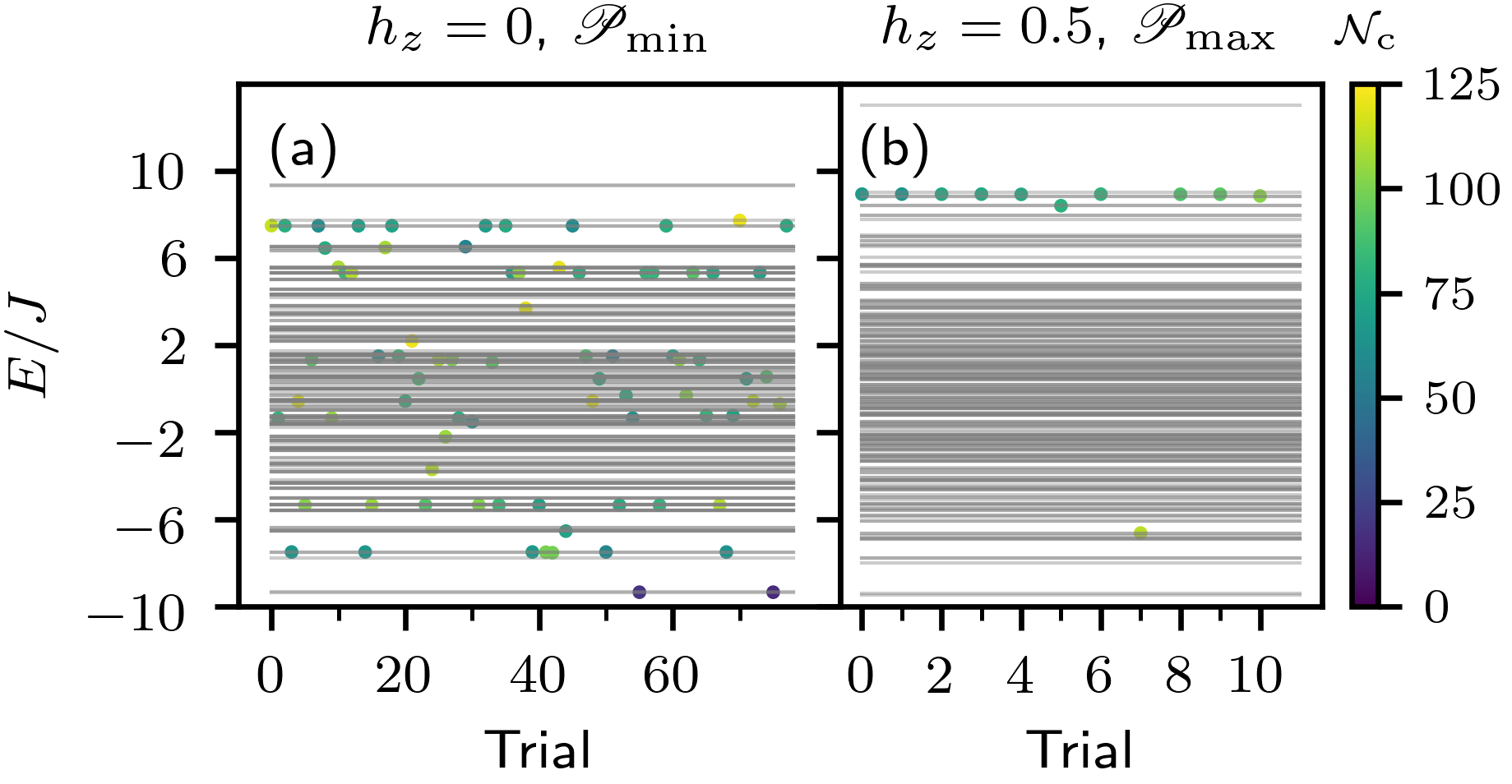}
    \caption{Adaptive VQE-X results at $N=8$ for (a) the integrable TFIM using $\mathscr{P}_{\rm min}$ and (b) the nonintegrable MFIM using $\mathscr{P}_{\rm max}$. The data points represent converged trials and are color coded according to $\mathcal N_{c}$. There is a marked difference in the number of converged trials between the integrable and nonintegrable cases.}
    \label{fig:N8}
\end{figure}

The observed exponential scaling of $\mathcal{N}_c$ with $N$ implies that classical optimization over exponentially many variational parameters is required. This is highly costly and poses the primary bottleneck for reaching larger system sizes with adaptive VQE-X. In Fig.~\ref{fig:func_eval}, we plot the number of cost function evaluations needed to complete the Nelder-Mead optimization as a function of the number of variational parameters during the early iterations of an implementation of adaptive VQE-X for the nonintegrable MFIM with $N=8$. The red curve is a quadratic fit to the data, indicating that the number of cost function evaluations grows roughly as the square of the number of parameters. As $\mathcal{N}_c$ scales exponentially with $N$, it takes $\sim$ 200 iterations and cumulatively $\sim 10^7$ cost function evaluations, each one of which is made of hundreds of matrix-vector multiplications, in order to converge to an eigenstate. To further demonstrate this, we compare in Fig.~\ref{fig:N8} the adaptive VQE-X results for the integrable TFIM and nonintegrable MFIM with $N=8$. Within $\sim$ 120 iterations, nearly 80 trials converged for the TFIM using $\mathscr{P}_{\rm min}$, with a wide coverage of the energy spectrum. On the other hand, only about 10 trials converged to a few eigenstates using $\mathscr{P}_{\rm max}$ for the MFIM.

Indeed, the challenge of excessively many variational parameters is likely to face any variational algorithm that attempts to approximate individual finite-energy-density eigenstates of generic local Hamiltonians. Indeed, generic quantum states, of which such highly excited eigenstates are believed to be representative examples, require exponentially many gates to prepare~\cite{Nielsen02}. Our results in Sec.~\ref{sec: Comparison to Adaptive Folded-Spectrum Method} corroborate this intuition; they show that the adaptive FSM approach yields circuits with a similar number of variational parameters on average as those produced within adaptive VQE-X. 
It would be interesting to see whether applying the same methodology to other cost functions capable of identifying arbitrary eigenstates, such as the ``eigenstate witness" proposed in Ref.~\cite{WAVES}, could potentially yield shorter circuits or an increased convergence rate relative to the cost functions explored here. We note, however, that this cost function is difficult to implement for many-body systems on NISQ devices, as it involves an ancilla qubit, a controlled-$e^{-iHt}$ operation, and full tomography of the ancilla. Additionally, an exponentially long evolution time $t$ is necessary to resolve the exponentially small gaps in the middle of the spectrum.
In any case, finding strategies to reduce the number of variational parameters---e.g., by modifying the operator pool or foregoing the adaptive strategy in favor of a fixed family of ans\"atze---will be crucial for making further progress on this problem.

While our results demonstrate that approximating individual highly excited eigenstates is a challenging task, they still provide some encouraging indications that quantum algorithms could be used to study highly excited states of quantum many-body systems. For example, Fig.~\ref{fig:1}(c) demonstrates that adaptive VQE-X can be used to reproduce averages over an energy window of quantities of interest such as the entanglement entropy. This suggests that one could potentially modify the adaptive VQE-X algorithm to provide states that serve as \textit{estimators} of microcanonical averages of observables and entanglement measures at finite energy density.  Such a strategy should be feasible in systems obeying the strong ETH, where expectation values of observables for any eigenstate in a fixed energy window are expected to match averages over a Gibbs ensemble at a corresponding temperature, up to corrections that vanish in the thermodynamic limit~\cite{Deutsch1991,Srednicki1994,Rigol2008,DAlessio2016,Deutsch18}. Whether such an estimator can be constructed using, e.g., a fixed variational ansatz with polynomially many parameters, such as the Hamiltonian variational ansatz of Ref.~\cite{Wiersema20}, is an interesting question for future work. Alternatively, converged adaptive VQE-X trials can potentially serve as a basis for computing microcanonical averages. For example, given a converged variational state $\ket{\Psi}$, one can construct a state $\ket{\Psi(\theta)}=e^{-i\theta\mathcal O}\ket{\Psi}$, where $\mathcal O$ is some (local) Hermitian operator. For sufficiently small $\theta$, the state $\ket{\Psi(\theta)}$ admixes $\ket{\Psi}$ with states nearby in energy. Taking expectation values of local operators in this state and averaging over $\theta$ and choices of operators $\mathcal{O}$ may allow for a better approximation of microcanonical averages than the state $\ket{\Psi}$ alone.

Another potential avenue for future work is to use (adaptive) VQE-X to study highly excited states of disordered quantum many-body systems. When disorder is sufficiently strong, such systems are believed to enter a many-body localized (MBL) phase~\cite{Abanin2019} in which emergent local conserved quantities render even highly excited states area-law entangled~\cite{Serbyn13,Bauer13,Nandkishore14}. The reduced entanglement of highly excited states in such systems may make them more amenable to representation by compact variational ans\"atze---indeed, this fact also enables classical simulation of such systems using tensor network approaches~\cite{Khemani16,Pollmann16}. Highly excited states with low entanglement also arise in models with quantum many-body scars~\cite{Turner2018,Moudgalya18,Schecter2019}. A variant of (adaptive) VQE-X (or the FSM) can be envisioned in which the cost function also contains an entanglement witness, such as the quantum Fisher information~\cite{Pezz2009}, such that the algorithm seeks to minimize both the entanglement and the energy variance of the variational state, potentially providing circuit representations of (rare) low-entanglement states in the spectrum. 

\textit{Note Added}: The data and source code used for this work have been made publicly available~\cite{code}.

\begin{acknowledgments}
This material is based upon work supported by the National Science Foundation under Grant No.~DMR-2038010 (T.I.~and P.P.O.). F.Z., N.G. and Y.Y. were supported by the U.S. Department of Energy (DOE), Office of Science, Basic Energy Sciences, Division of Materials Sciences and Engineering, and performed the research at the Ames Laboratory, which is operated for the U.S. DOE by Iowa State University under Contract DE-AC02-07CH11358.
\end{acknowledgments}

\bibliographystyle{apsrev4-2}
\bibliography{refs.bib}

\begin{thebibliography}{67}%
\makeatletter
\providecommand \@ifxundefined [1]{%
 \@ifx{#1\undefined}
}%
\providecommand \@ifnum [1]{%
 \ifnum #1\expandafter \@firstoftwo
 \else \expandafter \@secondoftwo
 \fi
}%
\providecommand \@ifx [1]{%
 \ifx #1\expandafter \@firstoftwo
 \else \expandafter \@secondoftwo
 \fi
}%
\providecommand \natexlab [1]{#1}%
\providecommand \enquote  [1]{``#1''}%
\providecommand \bibnamefont  [1]{#1}%
\providecommand \bibfnamefont [1]{#1}%
\providecommand \citenamefont [1]{#1}%
\providecommand \href@noop [0]{\@secondoftwo}%
\providecommand \href [0]{\begingroup \@sanitize@url \@href}%
\providecommand \@href[1]{\@@startlink{#1}\@@href}%
\providecommand \@@href[1]{\endgroup#1\@@endlink}%
\providecommand \@sanitize@url [0]{\catcode `\\12\catcode `\$12\catcode
  `\&12\catcode `\#12\catcode `\^12\catcode `\_12\catcode `\%12\relax}%
\providecommand \@@startlink[1]{}%
\providecommand \@@endlink[0]{}%
\providecommand \url  [0]{\begingroup\@sanitize@url \@url }%
\providecommand \@url [1]{\endgroup\@href {#1}{\urlprefix }}%
\providecommand \urlprefix  [0]{URL }%
\providecommand \Eprint [0]{\href }%
\providecommand \doibase [0]{https://doi.org/}%
\providecommand \selectlanguage [0]{\@gobble}%
\providecommand \bibinfo  [0]{\@secondoftwo}%
\providecommand \bibfield  [0]{\@secondoftwo}%
\providecommand \translation [1]{[#1]}%
\providecommand \BibitemOpen [0]{}%
\providecommand \bibitemStop [0]{}%
\providecommand \bibitemNoStop [0]{.\EOS\space}%
\providecommand \EOS [0]{\spacefactor3000\relax}%
\providecommand \BibitemShut  [1]{\csname bibitem#1\endcsname}%
\let\auto@bib@innerbib\@empty
\bibitem [{\citenamefont {Preskill}(2018)}]{preskillQuantumComputingNISQ2018}%
  \BibitemOpen
  \bibfield  {author} {\bibinfo {author} {\bibfnamefont {J.}~\bibnamefont
  {Preskill}},\ }\href {https://doi.org/10.22331/q-2018-08-06-79} {\bibfield
  {journal} {\bibinfo  {journal} {Quantum}\ }\textbf {\bibinfo {volume} {2}},\
  \bibinfo {pages} {79} (\bibinfo {year} {2018})}\BibitemShut {NoStop}%
\bibitem [{\citenamefont {Bharti}\ \emph {et~al.}(2021)\citenamefont {Bharti},
  \citenamefont {Cervera-Lierta}, \citenamefont {Kyaw}, \citenamefont {Haug},
  \citenamefont {Alperin-Lea}, \citenamefont {Anand}, \citenamefont {Degroote},
  \citenamefont {Heimonen}, \citenamefont {Kottmann}, \citenamefont {Menke},
  \citenamefont {Mok}, \citenamefont {Sim}, \citenamefont {Kwek},\ and\
  \citenamefont {Aspuru-Guzik}}]{Bharti21NISQReview}%
  \BibitemOpen
  \bibfield  {author} {\bibinfo {author} {\bibfnamefont {K.}~\bibnamefont
  {Bharti}}, \bibinfo {author} {\bibfnamefont {A.}~\bibnamefont
  {Cervera-Lierta}}, \bibinfo {author} {\bibfnamefont {T.~H.}\ \bibnamefont
  {Kyaw}}, \bibinfo {author} {\bibfnamefont {T.}~\bibnamefont {Haug}}, \bibinfo
  {author} {\bibfnamefont {S.}~\bibnamefont {Alperin-Lea}}, \bibinfo {author}
  {\bibfnamefont {A.}~\bibnamefont {Anand}}, \bibinfo {author} {\bibfnamefont
  {M.}~\bibnamefont {Degroote}}, \bibinfo {author} {\bibfnamefont
  {H.}~\bibnamefont {Heimonen}}, \bibinfo {author} {\bibfnamefont {J.~S.}\
  \bibnamefont {Kottmann}}, \bibinfo {author} {\bibfnamefont {T.}~\bibnamefont
  {Menke}}, \bibinfo {author} {\bibfnamefont {W.-K.}\ \bibnamefont {Mok}},
  \bibinfo {author} {\bibfnamefont {S.}~\bibnamefont {Sim}}, \bibinfo {author}
  {\bibfnamefont {L.-C.}\ \bibnamefont {Kwek}},\ and\ \bibinfo {author}
  {\bibfnamefont {A.}~\bibnamefont {Aspuru-Guzik}},\ }\href@noop {} {}
  (\bibinfo {year} {2021}),\ \Eprint {https://arxiv.org/abs/2101.08448}
  {arXiv:2101.08448 [quant-ph]} \BibitemShut {NoStop}%
\bibitem [{\citenamefont {Cerezo}\ \emph {et~al.}(2020)\citenamefont {Cerezo},
  \citenamefont {Arrasmith}, \citenamefont {Babbush}, \citenamefont {Benjamin},
  \citenamefont {Endo}, \citenamefont {Fujii}, \citenamefont {McClean},
  \citenamefont {Mitarai}, \citenamefont {Yuan}, \citenamefont {Cincio},\ and\
  \citenamefont {Coles}}]{cerezo2020variational}%
  \BibitemOpen
  \bibfield  {author} {\bibinfo {author} {\bibfnamefont {M.}~\bibnamefont
  {Cerezo}}, \bibinfo {author} {\bibfnamefont {A.}~\bibnamefont {Arrasmith}},
  \bibinfo {author} {\bibfnamefont {R.}~\bibnamefont {Babbush}}, \bibinfo
  {author} {\bibfnamefont {S.~C.}\ \bibnamefont {Benjamin}}, \bibinfo {author}
  {\bibfnamefont {S.}~\bibnamefont {Endo}}, \bibinfo {author} {\bibfnamefont
  {K.}~\bibnamefont {Fujii}}, \bibinfo {author} {\bibfnamefont {J.~R.}\
  \bibnamefont {McClean}}, \bibinfo {author} {\bibfnamefont {K.}~\bibnamefont
  {Mitarai}}, \bibinfo {author} {\bibfnamefont {X.}~\bibnamefont {Yuan}},
  \bibinfo {author} {\bibfnamefont {L.}~\bibnamefont {Cincio}},\ and\ \bibinfo
  {author} {\bibfnamefont {P.~J.}\ \bibnamefont {Coles}},\ }\href@noop {} {}
  (\bibinfo {year} {2020}),\ \Eprint {https://arxiv.org/abs/2012.09265}
  {arXiv:2012.09265 [quant-ph]} \BibitemShut {NoStop}%
\bibitem [{\citenamefont {Peruzzo}\ \emph {et~al.}(2014)\citenamefont
  {Peruzzo}, \citenamefont {McClean}, \citenamefont {Shadbolt}, \citenamefont
  {Yung}, \citenamefont {Zhou}, \citenamefont {Love}, \citenamefont
  {{Aspuru-Guzik}},\ and\ \citenamefont
  {O'Brien}}]{peruzzoVariationalEigenvalueSolver2014}%
  \BibitemOpen
  \bibfield  {author} {\bibinfo {author} {\bibfnamefont {A.}~\bibnamefont
  {Peruzzo}}, \bibinfo {author} {\bibfnamefont {J.}~\bibnamefont {McClean}},
  \bibinfo {author} {\bibfnamefont {P.}~\bibnamefont {Shadbolt}}, \bibinfo
  {author} {\bibfnamefont {M.-H.}\ \bibnamefont {Yung}}, \bibinfo {author}
  {\bibfnamefont {X.-Q.}\ \bibnamefont {Zhou}}, \bibinfo {author}
  {\bibfnamefont {P.~J.}\ \bibnamefont {Love}}, \bibinfo {author}
  {\bibfnamefont {A.}~\bibnamefont {{Aspuru-Guzik}}},\ and\ \bibinfo {author}
  {\bibfnamefont {J.~L.}\ \bibnamefont {O'Brien}},\ }\href
  {https://doi.org/10.1038/ncomms5213} {\bibfield  {journal} {\bibinfo
  {journal} {Nat. Comm.}\ }\textbf {\bibinfo {volume} {5}},\ \bibinfo {pages}
  {1} (\bibinfo {year} {2014})}\BibitemShut {NoStop}%
\bibitem [{\citenamefont {McClean}\ \emph {et~al.}(2016)\citenamefont
  {McClean}, \citenamefont {Romero}, \citenamefont {Babbush},\ and\
  \citenamefont {Aspuru-Guzik}}]{McClean16}%
  \BibitemOpen
  \bibfield  {author} {\bibinfo {author} {\bibfnamefont {J.~R.}\ \bibnamefont
  {McClean}}, \bibinfo {author} {\bibfnamefont {J.}~\bibnamefont {Romero}},
  \bibinfo {author} {\bibfnamefont {R.}~\bibnamefont {Babbush}},\ and\ \bibinfo
  {author} {\bibfnamefont {A.}~\bibnamefont {Aspuru-Guzik}},\ }\href
  {https://doi.org/10.1088/1367-2630/18/2/023023} {\bibfield  {journal}
  {\bibinfo  {journal} {New Journal of Physics}\ }\textbf {\bibinfo {volume}
  {18}},\ \bibinfo {pages} {023023} (\bibinfo {year} {2016})}\BibitemShut
  {NoStop}%
\bibitem [{\citenamefont {Kandala}\ \emph {et~al.}(2017)\citenamefont
  {Kandala}, \citenamefont {Mezzacapo}, \citenamefont {Temme}, \citenamefont
  {Takita}, \citenamefont {Brink}, \citenamefont {Chow},\ and\ \citenamefont
  {Gambetta}}]{kandalaHardwareefficientVariationalQuantum2017}%
  \BibitemOpen
  \bibfield  {author} {\bibinfo {author} {\bibfnamefont {A.}~\bibnamefont
  {Kandala}}, \bibinfo {author} {\bibfnamefont {A.}~\bibnamefont {Mezzacapo}},
  \bibinfo {author} {\bibfnamefont {K.}~\bibnamefont {Temme}}, \bibinfo
  {author} {\bibfnamefont {M.}~\bibnamefont {Takita}}, \bibinfo {author}
  {\bibfnamefont {M.}~\bibnamefont {Brink}}, \bibinfo {author} {\bibfnamefont
  {J.~M.}\ \bibnamefont {Chow}},\ and\ \bibinfo {author} {\bibfnamefont
  {J.~M.}\ \bibnamefont {Gambetta}},\ }\href
  {https://doi.org/10.1038/nature23879} {\bibfield  {journal} {\bibinfo
  {journal} {Nature}\ }\textbf {\bibinfo {volume} {549}},\ \bibinfo {pages}
  {242} (\bibinfo {year} {2017})}\BibitemShut {NoStop}%
\bibitem [{\citenamefont {O'Malley}\ \emph {et~al.}(2016)\citenamefont
  {O'Malley}, \citenamefont {Babbush}, \citenamefont {Kivlichan}, \citenamefont
  {Romero}, \citenamefont {McClean}, \citenamefont {Barends}, \citenamefont
  {Kelly}, \citenamefont {Roushan}, \citenamefont {Tranter}, \citenamefont
  {Ding}, \citenamefont {Campbell}, \citenamefont {Chen}, \citenamefont {Chen},
  \citenamefont {Chiaro}, \citenamefont {Dunsworth}, \citenamefont {Fowler},
  \citenamefont {Jeffrey}, \citenamefont {Lucero}, \citenamefont {Megrant},
  \citenamefont {Mutus}, \citenamefont {Neeley}, \citenamefont {Neill},
  \citenamefont {Quintana}, \citenamefont {Sank}, \citenamefont {Vainsencher},
  \citenamefont {Wenner}, \citenamefont {White}, \citenamefont {Coveney},
  \citenamefont {Love}, \citenamefont {Neven}, \citenamefont {{Aspuru-Guzik}},\
  and\ \citenamefont {Martinis}}]{omalleyScalableQuantumSimulation2016}%
  \BibitemOpen
  \bibfield  {author} {\bibinfo {author} {\bibfnamefont {P.~J.~J.}\
  \bibnamefont {O'Malley}}, \bibinfo {author} {\bibfnamefont {R.}~\bibnamefont
  {Babbush}}, \bibinfo {author} {\bibfnamefont {I.~D.}\ \bibnamefont
  {Kivlichan}}, \bibinfo {author} {\bibfnamefont {J.}~\bibnamefont {Romero}},
  \bibinfo {author} {\bibfnamefont {J.~R.}\ \bibnamefont {McClean}}, \bibinfo
  {author} {\bibfnamefont {R.}~\bibnamefont {Barends}}, \bibinfo {author}
  {\bibfnamefont {J.}~\bibnamefont {Kelly}}, \bibinfo {author} {\bibfnamefont
  {P.}~\bibnamefont {Roushan}}, \bibinfo {author} {\bibfnamefont
  {A.}~\bibnamefont {Tranter}}, \bibinfo {author} {\bibfnamefont
  {N.}~\bibnamefont {Ding}}, \bibinfo {author} {\bibfnamefont {B.}~\bibnamefont
  {Campbell}}, \bibinfo {author} {\bibfnamefont {Y.}~\bibnamefont {Chen}},
  \bibinfo {author} {\bibfnamefont {Z.}~\bibnamefont {Chen}}, \bibinfo {author}
  {\bibfnamefont {B.}~\bibnamefont {Chiaro}}, \bibinfo {author} {\bibfnamefont
  {A.}~\bibnamefont {Dunsworth}}, \bibinfo {author} {\bibfnamefont {A.~G.}\
  \bibnamefont {Fowler}}, \bibinfo {author} {\bibfnamefont {E.}~\bibnamefont
  {Jeffrey}}, \bibinfo {author} {\bibfnamefont {E.}~\bibnamefont {Lucero}},
  \bibinfo {author} {\bibfnamefont {A.}~\bibnamefont {Megrant}}, \bibinfo
  {author} {\bibfnamefont {J.~Y.}\ \bibnamefont {Mutus}}, \bibinfo {author}
  {\bibfnamefont {M.}~\bibnamefont {Neeley}}, \bibinfo {author} {\bibfnamefont
  {C.}~\bibnamefont {Neill}}, \bibinfo {author} {\bibfnamefont
  {C.}~\bibnamefont {Quintana}}, \bibinfo {author} {\bibfnamefont
  {D.}~\bibnamefont {Sank}}, \bibinfo {author} {\bibfnamefont {A.}~\bibnamefont
  {Vainsencher}}, \bibinfo {author} {\bibfnamefont {J.}~\bibnamefont {Wenner}},
  \bibinfo {author} {\bibfnamefont {T.~C.}\ \bibnamefont {White}}, \bibinfo
  {author} {\bibfnamefont {P.~V.}\ \bibnamefont {Coveney}}, \bibinfo {author}
  {\bibfnamefont {P.~J.}\ \bibnamefont {Love}}, \bibinfo {author}
  {\bibfnamefont {H.}~\bibnamefont {Neven}}, \bibinfo {author} {\bibfnamefont
  {A.}~\bibnamefont {{Aspuru-Guzik}}},\ and\ \bibinfo {author} {\bibfnamefont
  {J.~M.}\ \bibnamefont {Martinis}},\ }\href
  {https://doi.org/10.1103/PhysRevX.6.031007} {\bibfield  {journal} {\bibinfo
  {journal} {Phys. Rev. X}\ }\textbf {\bibinfo {volume} {6}},\ \bibinfo {pages}
  {031007} (\bibinfo {year} {2016})}\BibitemShut {NoStop}%
\bibitem [{\citenamefont {Colless}\ \emph {et~al.}(2018)\citenamefont
  {Colless}, \citenamefont {Ramasesh}, \citenamefont {Dahlen}, \citenamefont
  {Blok}, \citenamefont {{Kimchi-Schwartz}}, \citenamefont {McClean},
  \citenamefont {Carter}, \citenamefont {{de Jong}},\ and\ \citenamefont
  {Siddiqi}}]{collessComputationMolecularSpectra2018}%
  \BibitemOpen
  \bibfield  {author} {\bibinfo {author} {\bibfnamefont {J.~I.}\ \bibnamefont
  {Colless}}, \bibinfo {author} {\bibfnamefont {V.~V.}\ \bibnamefont
  {Ramasesh}}, \bibinfo {author} {\bibfnamefont {D.}~\bibnamefont {Dahlen}},
  \bibinfo {author} {\bibfnamefont {M.~S.}\ \bibnamefont {Blok}}, \bibinfo
  {author} {\bibfnamefont {M.~E.}\ \bibnamefont {{Kimchi-Schwartz}}}, \bibinfo
  {author} {\bibfnamefont {J.~R.}\ \bibnamefont {McClean}}, \bibinfo {author}
  {\bibfnamefont {J.}~\bibnamefont {Carter}}, \bibinfo {author} {\bibfnamefont
  {W.~A.}\ \bibnamefont {{de Jong}}},\ and\ \bibinfo {author} {\bibfnamefont
  {I.}~\bibnamefont {Siddiqi}},\ }\href
  {https://doi.org/10.1103/PhysRevX.8.011021} {\bibfield  {journal} {\bibinfo
  {journal} {Phys. Rev. X}\ }\textbf {\bibinfo {volume} {8}},\ \bibinfo {pages}
  {011021} (\bibinfo {year} {2018})}\BibitemShut {NoStop}%
\bibitem [{\citenamefont {Grimsley}\ \emph {et~al.}(2019)\citenamefont
  {Grimsley}, \citenamefont {Economou}, \citenamefont {Barnes},\ and\
  \citenamefont {Mayhall}}]{grimsleyAdaptiveVariationalAlgorithm2019}%
  \BibitemOpen
  \bibfield  {author} {\bibinfo {author} {\bibfnamefont {H.~R.}\ \bibnamefont
  {Grimsley}}, \bibinfo {author} {\bibfnamefont {S.~E.}\ \bibnamefont
  {Economou}}, \bibinfo {author} {\bibfnamefont {E.}~\bibnamefont {Barnes}},\
  and\ \bibinfo {author} {\bibfnamefont {N.~J.}\ \bibnamefont {Mayhall}},\
  }\href {https://doi.org/10.1038/s41467-019-10988-2} {\bibfield  {journal}
  {\bibinfo  {journal} {Nat. Commun.}\ }\textbf {\bibinfo {volume} {10}},\
  \bibinfo {pages} {3007} (\bibinfo {year} {2019})}\BibitemShut {NoStop}%
\bibitem [{\citenamefont {Wecker}\ \emph {et~al.}(2015)\citenamefont {Wecker},
  \citenamefont {Hastings},\ and\ \citenamefont
  {Troyer}}]{weckerProgressPracticalQuantum2015}%
  \BibitemOpen
  \bibfield  {author} {\bibinfo {author} {\bibfnamefont {D.}~\bibnamefont
  {Wecker}}, \bibinfo {author} {\bibfnamefont {M.~B.}\ \bibnamefont
  {Hastings}},\ and\ \bibinfo {author} {\bibfnamefont {M.}~\bibnamefont
  {Troyer}},\ }\href {https://doi.org/10.1103/PhysRevA.92.042303} {\bibfield
  {journal} {\bibinfo  {journal} {Phys. Rev. A}\ }\textbf {\bibinfo {volume}
  {92}},\ \bibinfo {pages} {042303} (\bibinfo {year} {2015})}\BibitemShut
  {NoStop}%
\bibitem [{\citenamefont {Ho}\ and\ \citenamefont
  {Hsieh}(2019)}]{hoEfficientVariationalSimulation2019}%
  \BibitemOpen
  \bibfield  {author} {\bibinfo {author} {\bibfnamefont {W.~W.}\ \bibnamefont
  {Ho}}\ and\ \bibinfo {author} {\bibfnamefont {T.~H.}\ \bibnamefont {Hsieh}},\
  }\href {https://doi.org/10.21468/SciPostPhys.6.3.029} {\bibfield  {journal}
  {\bibinfo  {journal} {SciPost Physics}\ }\textbf {\bibinfo {volume} {6}},\
  \bibinfo {pages} {029} (\bibinfo {year} {2019})}\BibitemShut {NoStop}%
\bibitem [{\citenamefont {Matos}\ \emph {et~al.}(2021)\citenamefont {Matos},
  \citenamefont {Johri},\ and\ \citenamefont {Papi\ifmmode~\acute{c}\else
  \'{c}\fi{}}}]{Matos21}%
  \BibitemOpen
  \bibfield  {author} {\bibinfo {author} {\bibfnamefont {G.}~\bibnamefont
  {Matos}}, \bibinfo {author} {\bibfnamefont {S.}~\bibnamefont {Johri}},\ and\
  \bibinfo {author} {\bibfnamefont {Z.}~\bibnamefont
  {Papi\ifmmode~\acute{c}\else \'{c}\fi{}}},\ }\href
  {https://doi.org/10.1103/PRXQuantum.2.010309} {\bibfield  {journal} {\bibinfo
   {journal} {PRX Quantum}\ }\textbf {\bibinfo {volume} {2}},\ \bibinfo {pages}
  {010309} (\bibinfo {year} {2021})}\BibitemShut {NoStop}%
\bibitem [{\citenamefont {Klco}\ \emph {et~al.}(2018)\citenamefont {Klco},
  \citenamefont {Dumitrescu}, \citenamefont {McCaskey}, \citenamefont {Morris},
  \citenamefont {Pooser}, \citenamefont {Sanz}, \citenamefont {Solano},
  \citenamefont {Lougovski},\ and\ \citenamefont {Savage}}]{Klco18}%
  \BibitemOpen
  \bibfield  {author} {\bibinfo {author} {\bibfnamefont {N.}~\bibnamefont
  {Klco}}, \bibinfo {author} {\bibfnamefont {E.~F.}\ \bibnamefont
  {Dumitrescu}}, \bibinfo {author} {\bibfnamefont {A.~J.}\ \bibnamefont
  {McCaskey}}, \bibinfo {author} {\bibfnamefont {T.~D.}\ \bibnamefont
  {Morris}}, \bibinfo {author} {\bibfnamefont {R.~C.}\ \bibnamefont {Pooser}},
  \bibinfo {author} {\bibfnamefont {M.}~\bibnamefont {Sanz}}, \bibinfo {author}
  {\bibfnamefont {E.}~\bibnamefont {Solano}}, \bibinfo {author} {\bibfnamefont
  {P.}~\bibnamefont {Lougovski}},\ and\ \bibinfo {author} {\bibfnamefont
  {M.~J.}\ \bibnamefont {Savage}},\ }\href
  {https://doi.org/10.1103/PhysRevA.98.032331} {\bibfield  {journal} {\bibinfo
  {journal} {Phys. Rev. A}\ }\textbf {\bibinfo {volume} {98}},\ \bibinfo
  {pages} {032331} (\bibinfo {year} {2018})}\BibitemShut {NoStop}%
\bibitem [{\citenamefont {Kokail}\ \emph {et~al.}(2019)\citenamefont {Kokail},
  \citenamefont {Maier}, \citenamefont {van Bijnen}, \citenamefont {Brydges},
  \citenamefont {Joshi}, \citenamefont {Jurcevic}, \citenamefont {Muschik},
  \citenamefont {Silvi}, \citenamefont {Blatt}, \citenamefont {Roos} \emph
  {et~al.}}]{kokail2019self}%
  \BibitemOpen
  \bibfield  {author} {\bibinfo {author} {\bibfnamefont {C.}~\bibnamefont
  {Kokail}}, \bibinfo {author} {\bibfnamefont {C.}~\bibnamefont {Maier}},
  \bibinfo {author} {\bibfnamefont {R.}~\bibnamefont {van Bijnen}}, \bibinfo
  {author} {\bibfnamefont {T.}~\bibnamefont {Brydges}}, \bibinfo {author}
  {\bibfnamefont {M.~K.}\ \bibnamefont {Joshi}}, \bibinfo {author}
  {\bibfnamefont {P.}~\bibnamefont {Jurcevic}}, \bibinfo {author}
  {\bibfnamefont {C.~A.}\ \bibnamefont {Muschik}}, \bibinfo {author}
  {\bibfnamefont {P.}~\bibnamefont {Silvi}}, \bibinfo {author} {\bibfnamefont
  {R.}~\bibnamefont {Blatt}}, \bibinfo {author} {\bibfnamefont {C.~F.}\
  \bibnamefont {Roos}}, \emph {et~al.},\ }\href
  {https://www.nature.com/articles/s41586-019-1177-4} {\bibfield  {journal}
  {\bibinfo  {journal} {Nature}\ }\textbf {\bibinfo {volume} {569}},\ \bibinfo
  {pages} {355} (\bibinfo {year} {2019})}\BibitemShut {NoStop}%
\bibitem [{\citenamefont {McClean}\ \emph {et~al.}(2017)\citenamefont
  {McClean}, \citenamefont {{Kimchi-Schwartz}}, \citenamefont {Carter},\ and\
  \citenamefont {{de Jong}}}]{mccleanHybridQuantumclassicalHierarchy2017}%
  \BibitemOpen
  \bibfield  {author} {\bibinfo {author} {\bibfnamefont {J.~R.}\ \bibnamefont
  {McClean}}, \bibinfo {author} {\bibfnamefont {M.~E.}\ \bibnamefont
  {{Kimchi-Schwartz}}}, \bibinfo {author} {\bibfnamefont {J.}~\bibnamefont
  {Carter}},\ and\ \bibinfo {author} {\bibfnamefont {W.~A.}\ \bibnamefont {{de
  Jong}}},\ }\href {https://doi.org/10.1103/PhysRevA.95.042308} {\bibfield
  {journal} {\bibinfo  {journal} {Phys. Rev. A}\ }\textbf {\bibinfo {volume}
  {95}},\ \bibinfo {pages} {042308} (\bibinfo {year} {2017})}\BibitemShut
  {NoStop}%
\bibitem [{\citenamefont {Stair}\ \emph {et~al.}(2020)\citenamefont {Stair},
  \citenamefont {Huang},\ and\ \citenamefont
  {Evangelista}}]{stairMultireferenceQuantumKrylov2020}%
  \BibitemOpen
  \bibfield  {author} {\bibinfo {author} {\bibfnamefont {N.~H.}\ \bibnamefont
  {Stair}}, \bibinfo {author} {\bibfnamefont {R.}~\bibnamefont {Huang}},\ and\
  \bibinfo {author} {\bibfnamefont {F.~A.}\ \bibnamefont {Evangelista}},\
  }\href {https://doi.org/10.1021/acs.jctc.9b01125} {\bibfield  {journal}
  {\bibinfo  {journal} {J. Chem. Theory Comput.}\ }\textbf {\bibinfo {volume}
  {16}},\ \bibinfo {pages} {2236} (\bibinfo {year} {2020})}\BibitemShut
  {NoStop}%
\bibitem [{\citenamefont {Nakanishi}\ \emph {et~al.}(2019)\citenamefont
  {Nakanishi}, \citenamefont {Mitarai},\ and\ \citenamefont
  {Fujii}}]{nakanishiSubspacesearchVariationalQuantum2019}%
  \BibitemOpen
  \bibfield  {author} {\bibinfo {author} {\bibfnamefont {K.~M.}\ \bibnamefont
  {Nakanishi}}, \bibinfo {author} {\bibfnamefont {K.}~\bibnamefont {Mitarai}},\
  and\ \bibinfo {author} {\bibfnamefont {K.}~\bibnamefont {Fujii}},\ }\href
  {https://doi.org/10.1103/PhysRevResearch.1.033062} {\bibfield  {journal}
  {\bibinfo  {journal} {Phys. Rev. Research}\ }\textbf {\bibinfo {volume}
  {1}},\ \bibinfo {pages} {033062} (\bibinfo {year} {2019})}\BibitemShut
  {NoStop}%
\bibitem [{\citenamefont {Parrish}\ \emph {et~al.}(2019)\citenamefont
  {Parrish}, \citenamefont {Hohenstein}, \citenamefont {McMahon},\ and\
  \citenamefont {Mart{\'i}nez}}]{parrishQuantumComputationElectronic2019}%
  \BibitemOpen
  \bibfield  {author} {\bibinfo {author} {\bibfnamefont {R.~M.}\ \bibnamefont
  {Parrish}}, \bibinfo {author} {\bibfnamefont {E.~G.}\ \bibnamefont
  {Hohenstein}}, \bibinfo {author} {\bibfnamefont {P.~L.}\ \bibnamefont
  {McMahon}},\ and\ \bibinfo {author} {\bibfnamefont {T.~J.}\ \bibnamefont
  {Mart{\'i}nez}},\ }\href {https://doi.org/10.1103/PhysRevLett.122.230401}
  {\bibfield  {journal} {\bibinfo  {journal} {Phys. Rev. Lett.}\ }\textbf
  {\bibinfo {volume} {122}},\ \bibinfo {pages} {230401} (\bibinfo {year}
  {2019})}\BibitemShut {NoStop}%
\bibitem [{\citenamefont {Higgott}\ \emph {et~al.}(2019)\citenamefont
  {Higgott}, \citenamefont {Wang},\ and\ \citenamefont
  {Brierley}}]{higgottVariationalQuantumComputation2019}%
  \BibitemOpen
  \bibfield  {author} {\bibinfo {author} {\bibfnamefont {O.}~\bibnamefont
  {Higgott}}, \bibinfo {author} {\bibfnamefont {D.}~\bibnamefont {Wang}},\ and\
  \bibinfo {author} {\bibfnamefont {S.}~\bibnamefont {Brierley}},\ }\href
  {https://doi.org/10.22331/q-2019-07-01-156} {\bibfield  {journal} {\bibinfo
  {journal} {Quantum}\ }\textbf {\bibinfo {volume} {3}},\ \bibinfo {pages}
  {156} (\bibinfo {year} {2019})}\BibitemShut {NoStop}%
\bibitem [{\citenamefont {Hastings}(2007)}]{hastings2007area}%
  \BibitemOpen
  \bibfield  {author} {\bibinfo {author} {\bibfnamefont {M.~B.}\ \bibnamefont
  {Hastings}},\ }\href
  {https://iopscience.iop.org/article/10.1088/1742-5468/2007/08/P08024}
  {\bibfield  {journal} {\bibinfo  {journal} {J Stat. Mech.: Theory Exp.}\
  }\textbf {\bibinfo {volume} {2007}},\ \bibinfo {pages} {P08024} (\bibinfo
  {year} {2007})}\BibitemShut {NoStop}%
\bibitem [{\citenamefont {White}(1992)}]{White92}%
  \BibitemOpen
  \bibfield  {author} {\bibinfo {author} {\bibfnamefont {S.~R.}\ \bibnamefont
  {White}},\ }\href {https://doi.org/10.1103/PhysRevLett.69.2863} {\bibfield
  {journal} {\bibinfo  {journal} {Phys. Rev. Lett.}\ }\textbf {\bibinfo
  {volume} {69}},\ \bibinfo {pages} {2863} (\bibinfo {year}
  {1992})}\BibitemShut {NoStop}%
\bibitem [{\citenamefont
  {Schollw{\"o}ck}(2011)}]{schollwockDensitymatrixRenormalizationGroup2011}%
  \BibitemOpen
  \bibfield  {author} {\bibinfo {author} {\bibfnamefont {U.}~\bibnamefont
  {Schollw{\"o}ck}},\ }\href {https://doi.org/10.1016/j.aop.2010.09.012}
  {\bibfield  {journal} {\bibinfo  {journal} {Ann. Phys.}\ }\bibinfo {series}
  {January 2011 {{Special Issue}}},\ \textbf {\bibinfo {volume} {326}},\
  \bibinfo {pages} {96} (\bibinfo {year} {2011})}\BibitemShut {NoStop}%
\bibitem [{\citenamefont {Deutsch}(1991)}]{Deutsch1991}%
  \BibitemOpen
  \bibfield  {author} {\bibinfo {author} {\bibfnamefont {J.~M.}\ \bibnamefont
  {Deutsch}},\ }\href {https://doi.org/10.1103/physreva.43.2046} {\bibfield
  {journal} {\bibinfo  {journal} {Phys. Rev. A}\ }\textbf {\bibinfo {volume}
  {43}},\ \bibinfo {pages} {2046} (\bibinfo {year} {1991})}\BibitemShut
  {NoStop}%
\bibitem [{\citenamefont {Srednicki}(1994)}]{Srednicki1994}%
  \BibitemOpen
  \bibfield  {author} {\bibinfo {author} {\bibfnamefont {M.}~\bibnamefont
  {Srednicki}},\ }\href {https://doi.org/10.1103/physreve.50.888} {\bibfield
  {journal} {\bibinfo  {journal} {Phys. Rev. E}\ }\textbf {\bibinfo {volume}
  {50}},\ \bibinfo {pages} {888} (\bibinfo {year} {1994})}\BibitemShut
  {NoStop}%
\bibitem [{\citenamefont {Rigol}\ \emph {et~al.}(2008)\citenamefont {Rigol},
  \citenamefont {Dunjko},\ and\ \citenamefont {Olshanii}}]{Rigol2008}%
  \BibitemOpen
  \bibfield  {author} {\bibinfo {author} {\bibfnamefont {M.}~\bibnamefont
  {Rigol}}, \bibinfo {author} {\bibfnamefont {V.}~\bibnamefont {Dunjko}},\ and\
  \bibinfo {author} {\bibfnamefont {M.}~\bibnamefont {Olshanii}},\ }\href
  {https://doi.org/10.1038/nature06838} {\bibfield  {journal} {\bibinfo
  {journal} {Nature}\ }\textbf {\bibinfo {volume} {452}},\ \bibinfo {pages}
  {854} (\bibinfo {year} {2008})}\BibitemShut {NoStop}%
\bibitem [{\citenamefont {D'Alessio}\ \emph {et~al.}(2016)\citenamefont
  {D'Alessio}, \citenamefont {Kafri}, \citenamefont {Polkovnikov},\ and\
  \citenamefont {Rigol}}]{DAlessio2016}%
  \BibitemOpen
  \bibfield  {author} {\bibinfo {author} {\bibfnamefont {L.}~\bibnamefont
  {D'Alessio}}, \bibinfo {author} {\bibfnamefont {Y.}~\bibnamefont {Kafri}},
  \bibinfo {author} {\bibfnamefont {A.}~\bibnamefont {Polkovnikov}},\ and\
  \bibinfo {author} {\bibfnamefont {M.}~\bibnamefont {Rigol}},\ }\href
  {https://doi.org/10.1080/00018732.2016.1198134} {\bibfield  {journal}
  {\bibinfo  {journal} {Adv. Phys.}\ }\textbf {\bibinfo {volume} {65}},\
  \bibinfo {pages} {239} (\bibinfo {year} {2016})}\BibitemShut {NoStop}%
\bibitem [{\citenamefont {Deutsch}(2018)}]{Deutsch18}%
  \BibitemOpen
  \bibfield  {author} {\bibinfo {author} {\bibfnamefont {J.~M.}\ \bibnamefont
  {Deutsch}},\ }\href
  {https://iopscience.iop.org/article/10.1088/1361-6633/aac9f1} {\bibfield
  {journal} {\bibinfo  {journal} {Reports on Progress in Physics}\ }\textbf
  {\bibinfo {volume} {81}},\ \bibinfo {pages} {082001} (\bibinfo {year}
  {2018})}\BibitemShut {NoStop}%
\bibitem [{\citenamefont {Deutsch}(2010)}]{Deutsch10}%
  \BibitemOpen
  \bibfield  {author} {\bibinfo {author} {\bibfnamefont {J.}~\bibnamefont
  {Deutsch}},\ }\href
  {https://iopscience.iop.org/article/10.1088/1367-2630/12/7/075021} {\bibfield
   {journal} {\bibinfo  {journal} {New Journal of Physics}\ }\textbf {\bibinfo
  {volume} {12}},\ \bibinfo {pages} {075021} (\bibinfo {year}
  {2010})}\BibitemShut {NoStop}%
\bibitem [{\citenamefont {Garrison}\ and\ \citenamefont
  {Grover}(2018)}]{Garrison18}%
  \BibitemOpen
  \bibfield  {author} {\bibinfo {author} {\bibfnamefont {J.~R.}\ \bibnamefont
  {Garrison}}\ and\ \bibinfo {author} {\bibfnamefont {T.}~\bibnamefont
  {Grover}},\ }\href {https://doi.org/10.1103/PhysRevX.8.021026} {\bibfield
  {journal} {\bibinfo  {journal} {Phys. Rev. X}\ }\textbf {\bibinfo {volume}
  {8}},\ \bibinfo {pages} {021026} (\bibinfo {year} {2018})}\BibitemShut
  {NoStop}%
\bibitem [{\citenamefont {Page}(1993)}]{pageAverageEntropySubsystem1993}%
  \BibitemOpen
  \bibfield  {author} {\bibinfo {author} {\bibfnamefont {D.~N.}\ \bibnamefont
  {Page}},\ }\href {https://doi.org/10.1103/PhysRevLett.71.1291} {\bibfield
  {journal} {\bibinfo  {journal} {Phys. Rev. Lett.}\ }\textbf {\bibinfo
  {volume} {71}},\ \bibinfo {pages} {1291} (\bibinfo {year}
  {1993})}\BibitemShut {NoStop}%
\bibitem [{\citenamefont {Luitz}\ \emph {et~al.}(2015)\citenamefont {Luitz},
  \citenamefont {Laflorencie},\ and\ \citenamefont {Alet}}]{Luitz2015}%
  \BibitemOpen
  \bibfield  {author} {\bibinfo {author} {\bibfnamefont {D.~J.}\ \bibnamefont
  {Luitz}}, \bibinfo {author} {\bibfnamefont {N.}~\bibnamefont {Laflorencie}},\
  and\ \bibinfo {author} {\bibfnamefont {F.}~\bibnamefont {Alet}},\ }\href
  {https://doi.org/10.1103/physrevb.91.081103} {\bibfield  {journal} {\bibinfo
  {journal} {Phys. Rev. B}\ }\textbf {\bibinfo {volume} {91}},\ \bibinfo
  {pages} {081103} (\bibinfo {year} {2015})}\BibitemShut {NoStop}%
\bibitem [{\citenamefont {Naldesi}\ \emph {et~al.}(2016)\citenamefont
  {Naldesi}, \citenamefont {Ercolessi},\ and\ \citenamefont
  {Roscilde}}]{Naldesi16}%
  \BibitemOpen
  \bibfield  {author} {\bibinfo {author} {\bibfnamefont {P.}~\bibnamefont
  {Naldesi}}, \bibinfo {author} {\bibfnamefont {E.}~\bibnamefont {Ercolessi}},\
  and\ \bibinfo {author} {\bibfnamefont {T.}~\bibnamefont {Roscilde}},\ }\href
  {https://doi.org/10.21468/SciPostPhys.1.1.010} {\bibfield  {journal}
  {\bibinfo  {journal} {SciPost Phys.}\ }\textbf {\bibinfo {volume} {1}},\
  \bibinfo {pages} {010} (\bibinfo {year} {2016})}\BibitemShut {NoStop}%
\bibitem [{\citenamefont {Nag}\ and\ \citenamefont {Garg}(2017)}]{Nag17}%
  \BibitemOpen
  \bibfield  {author} {\bibinfo {author} {\bibfnamefont {S.}~\bibnamefont
  {Nag}}\ and\ \bibinfo {author} {\bibfnamefont {A.}~\bibnamefont {Garg}},\
  }\href {https://doi.org/10.1103/PhysRevB.96.060203} {\bibfield  {journal}
  {\bibinfo  {journal} {Phys. Rev. B}\ }\textbf {\bibinfo {volume} {96}},\
  \bibinfo {pages} {060203} (\bibinfo {year} {2017})}\BibitemShut {NoStop}%
\bibitem [{\citenamefont {Wei}\ \emph {et~al.}(2019)\citenamefont {Wei},
  \citenamefont {Cheng}, \citenamefont {Xianlong},\ and\ \citenamefont
  {Mondaini}}]{Wei19}%
  \BibitemOpen
  \bibfield  {author} {\bibinfo {author} {\bibfnamefont {X.}~\bibnamefont
  {Wei}}, \bibinfo {author} {\bibfnamefont {C.}~\bibnamefont {Cheng}}, \bibinfo
  {author} {\bibfnamefont {G.}~\bibnamefont {Xianlong}},\ and\ \bibinfo
  {author} {\bibfnamefont {R.}~\bibnamefont {Mondaini}},\ }\href
  {https://doi.org/10.1103/physrevb.99.165137} {\bibfield  {journal} {\bibinfo
  {journal} {Physical Review B}\ }\textbf {\bibinfo {volume} {99}},\ \bibinfo
  {pages} {165137} (\bibinfo {year} {2019})}\BibitemShut {NoStop}%
\bibitem [{\citenamefont {Zhang}\ \emph
  {et~al.}(2021{\natexlab{a}})\citenamefont {Zhang}, \citenamefont {Ke},
  \citenamefont {Liu},\ and\ \citenamefont {Lee}}]{Zhang21}%
  \BibitemOpen
  \bibfield  {author} {\bibinfo {author} {\bibfnamefont {L.}~\bibnamefont
  {Zhang}}, \bibinfo {author} {\bibfnamefont {Y.}~\bibnamefont {Ke}}, \bibinfo
  {author} {\bibfnamefont {W.}~\bibnamefont {Liu}},\ and\ \bibinfo {author}
  {\bibfnamefont {C.}~\bibnamefont {Lee}},\ }\href
  {https://doi.org/10.1103/PhysRevA.103.023323} {\bibfield  {journal} {\bibinfo
   {journal} {Phys. Rev. A}\ }\textbf {\bibinfo {volume} {103}},\ \bibinfo
  {pages} {023323} (\bibinfo {year} {2021}{\natexlab{a}})}\BibitemShut
  {NoStop}%
\bibitem [{\citenamefont {Schecter}\ \emph {et~al.}(2018)\citenamefont
  {Schecter}, \citenamefont {Iadecola},\ and\ \citenamefont
  {{Das~Sarma}}}]{Schecter2018}%
  \BibitemOpen
  \bibfield  {author} {\bibinfo {author} {\bibfnamefont {M.}~\bibnamefont
  {Schecter}}, \bibinfo {author} {\bibfnamefont {T.}~\bibnamefont {Iadecola}},\
  and\ \bibinfo {author} {\bibfnamefont {S.}~\bibnamefont {{Das~Sarma}}},\
  }\href {https://doi.org/10.1103/physrevb.98.174201} {\bibfield  {journal}
  {\bibinfo  {journal} {Phys. Rev. B}\ }\textbf {\bibinfo {volume} {98}},\
  \bibinfo {pages} {174201} (\bibinfo {year} {2018})}\BibitemShut {NoStop}%
\bibitem [{\citenamefont {{De~Roeck}}\ \emph {et~al.}(2016)\citenamefont
  {{De~Roeck}}, \citenamefont {Huveneers}, \citenamefont {M\"{u}ller},\ and\
  \citenamefont {Schiulaz}}]{DeRoeck2016}%
  \BibitemOpen
  \bibfield  {author} {\bibinfo {author} {\bibfnamefont {W.}~\bibnamefont
  {{De~Roeck}}}, \bibinfo {author} {\bibfnamefont {F.}~\bibnamefont
  {Huveneers}}, \bibinfo {author} {\bibfnamefont {M.}~\bibnamefont
  {M\"{u}ller}},\ and\ \bibinfo {author} {\bibfnamefont {M.}~\bibnamefont
  {Schiulaz}},\ }\href {https://doi.org/10.1103/physrevb.93.014203} {\bibfield
  {journal} {\bibinfo  {journal} {Phys. Rev. B}\ }\textbf {\bibinfo {volume}
  {93}},\ \bibinfo {pages} {014203} (\bibinfo {year} {2016})}\BibitemShut
  {NoStop}%
\bibitem [{\citenamefont {Crowley}\ and\ \citenamefont
  {Chandran}(2020)}]{Crowley20}%
  \BibitemOpen
  \bibfield  {author} {\bibinfo {author} {\bibfnamefont {P.~J.~D.}\
  \bibnamefont {Crowley}}\ and\ \bibinfo {author} {\bibfnamefont
  {A.}~\bibnamefont {Chandran}},\ }\href
  {https://doi.org/10.1103/PhysRevResearch.2.033262} {\bibfield  {journal}
  {\bibinfo  {journal} {Phys. Rev. Research}\ }\textbf {\bibinfo {volume}
  {2}},\ \bibinfo {pages} {033262} (\bibinfo {year} {2020})}\BibitemShut
  {NoStop}%
\bibitem [{\citenamefont {Brighi}\ \emph {et~al.}(2020)\citenamefont {Brighi},
  \citenamefont {Abanin},\ and\ \citenamefont {Serbyn}}]{Brighi20}%
  \BibitemOpen
  \bibfield  {author} {\bibinfo {author} {\bibfnamefont {P.}~\bibnamefont
  {Brighi}}, \bibinfo {author} {\bibfnamefont {D.~A.}\ \bibnamefont {Abanin}},\
  and\ \bibinfo {author} {\bibfnamefont {M.}~\bibnamefont {Serbyn}},\ }\href
  {https://doi.org/10.1103/PhysRevB.102.060202} {\bibfield  {journal} {\bibinfo
   {journal} {Phys. Rev. B}\ }\textbf {\bibinfo {volume} {102}},\ \bibinfo
  {pages} {060202} (\bibinfo {year} {2020})}\BibitemShut {NoStop}%
\bibitem [{\citenamefont {Sala}\ \emph {et~al.}(2020)\citenamefont {Sala},
  \citenamefont {Rakovszky}, \citenamefont {Verresen}, \citenamefont {Knap},\
  and\ \citenamefont {Pollmann}}]{Sala19}%
  \BibitemOpen
  \bibfield  {author} {\bibinfo {author} {\bibfnamefont {P.}~\bibnamefont
  {Sala}}, \bibinfo {author} {\bibfnamefont {T.}~\bibnamefont {Rakovszky}},
  \bibinfo {author} {\bibfnamefont {R.}~\bibnamefont {Verresen}}, \bibinfo
  {author} {\bibfnamefont {M.}~\bibnamefont {Knap}},\ and\ \bibinfo {author}
  {\bibfnamefont {F.}~\bibnamefont {Pollmann}},\ }\href
  {https://doi.org/10.1103/PhysRevX.10.011047} {\bibfield  {journal} {\bibinfo
  {journal} {Phys. Rev. X}\ }\textbf {\bibinfo {volume} {10}},\ \bibinfo
  {pages} {011047} (\bibinfo {year} {2020})}\BibitemShut {NoStop}%
\bibitem [{\citenamefont {Khemani}\ \emph {et~al.}(2020)\citenamefont
  {Khemani}, \citenamefont {Hermele},\ and\ \citenamefont
  {Nandkishore}}]{Khemani19}%
  \BibitemOpen
  \bibfield  {author} {\bibinfo {author} {\bibfnamefont {V.}~\bibnamefont
  {Khemani}}, \bibinfo {author} {\bibfnamefont {M.}~\bibnamefont {Hermele}},\
  and\ \bibinfo {author} {\bibfnamefont {R.}~\bibnamefont {Nandkishore}},\
  }\href {https://doi.org/10.1103/PhysRevB.101.174204} {\bibfield  {journal}
  {\bibinfo  {journal} {Phys. Rev. B}\ }\textbf {\bibinfo {volume} {101}},\
  \bibinfo {pages} {174204} (\bibinfo {year} {2020})}\BibitemShut {NoStop}%
\bibitem [{\citenamefont {Turner}\ \emph {et~al.}(2018)\citenamefont {Turner},
  \citenamefont {Michailidis}, \citenamefont {Abanin}, \citenamefont {Serbyn},\
  and\ \citenamefont {Papi{\'{c}}}}]{Turner2018}%
  \BibitemOpen
  \bibfield  {author} {\bibinfo {author} {\bibfnamefont {C.~J.}\ \bibnamefont
  {Turner}}, \bibinfo {author} {\bibfnamefont {A.~A.}\ \bibnamefont
  {Michailidis}}, \bibinfo {author} {\bibfnamefont {D.~A.}\ \bibnamefont
  {Abanin}}, \bibinfo {author} {\bibfnamefont {M.}~\bibnamefont {Serbyn}},\
  and\ \bibinfo {author} {\bibfnamefont {Z.}~\bibnamefont {Papi{\'{c}}}},\
  }\href {https://doi.org/10.1038/s41567-018-0137-5} {\bibfield  {journal}
  {\bibinfo  {journal} {Nat. Phys.}\ }\textbf {\bibinfo {volume} {14}},\
  \bibinfo {pages} {745} (\bibinfo {year} {2018})}\BibitemShut {NoStop}%
\bibitem [{\citenamefont {Moudgalya}\ \emph {et~al.}(2018)\citenamefont
  {Moudgalya}, \citenamefont {Regnault},\ and\ \citenamefont
  {Bernevig}}]{Moudgalya18}%
  \BibitemOpen
  \bibfield  {author} {\bibinfo {author} {\bibfnamefont {S.}~\bibnamefont
  {Moudgalya}}, \bibinfo {author} {\bibfnamefont {N.}~\bibnamefont
  {Regnault}},\ and\ \bibinfo {author} {\bibfnamefont {B.~A.}\ \bibnamefont
  {Bernevig}},\ }\href {https://doi.org/10.1103/PhysRevB.98.235156} {\bibfield
  {journal} {\bibinfo  {journal} {Phys. Rev. B}\ }\textbf {\bibinfo {volume}
  {98}},\ \bibinfo {pages} {235156} (\bibinfo {year} {2018})}\BibitemShut
  {NoStop}%
\bibitem [{\citenamefont {Schecter}\ and\ \citenamefont
  {Iadecola}(2019)}]{Schecter2019}%
  \BibitemOpen
  \bibfield  {author} {\bibinfo {author} {\bibfnamefont {M.}~\bibnamefont
  {Schecter}}\ and\ \bibinfo {author} {\bibfnamefont {T.}~\bibnamefont
  {Iadecola}},\ }\href {https://doi.org/10.1103/physrevlett.123.147201}
  {\bibfield  {journal} {\bibinfo  {journal} {Phys. Rev. Lett.}\ }\textbf
  {\bibinfo {volume} {123}},\ \bibinfo {pages} {147201} (\bibinfo {year}
  {2019})}\BibitemShut {NoStop}%
\bibitem [{\citenamefont {MacDonald}(1934)}]{macdonald1934modified}%
  \BibitemOpen
  \bibfield  {author} {\bibinfo {author} {\bibfnamefont {J.}~\bibnamefont
  {MacDonald}},\ }\href
  {https://journals.aps.org/pr/abstract/10.1103/PhysRev.46.828} {\bibfield
  {journal} {\bibinfo  {journal} {Phys. Rev.}\ }\textbf {\bibinfo {volume}
  {46}},\ \bibinfo {pages} {828} (\bibinfo {year} {1934})}\BibitemShut
  {NoStop}%
\bibitem [{\citenamefont {Wang}\ and\ \citenamefont
  {Zunger}(1994)}]{wang1994solving}%
  \BibitemOpen
  \bibfield  {author} {\bibinfo {author} {\bibfnamefont {L.-W.}\ \bibnamefont
  {Wang}}\ and\ \bibinfo {author} {\bibfnamefont {A.}~\bibnamefont {Zunger}},\
  }\href {https://aip.scitation.org/doi/abs/10.1063/1.466486} {\bibfield
  {journal} {\bibinfo  {journal} {J. Comp. Phys.}\ }\textbf {\bibinfo {volume}
  {100}},\ \bibinfo {pages} {2394} (\bibinfo {year} {1994})}\BibitemShut
  {NoStop}%
\bibitem [{\citenamefont {Santagati}\ \emph {et~al.}(2018)\citenamefont
  {Santagati}, \citenamefont {Wang}, \citenamefont {Gentile}, \citenamefont
  {Paesani}, \citenamefont {Wiebe}, \citenamefont {McClean}, \citenamefont
  {Morley-Short}, \citenamefont {Shadbolt}, \citenamefont {Bonneau},
  \citenamefont {Silverstone}, \citenamefont {Tew}, \citenamefont {Zhou},
  \citenamefont {O'Brien},\ and\ \citenamefont {Thompson}}]{WAVES}%
  \BibitemOpen
  \bibfield  {author} {\bibinfo {author} {\bibfnamefont {R.}~\bibnamefont
  {Santagati}}, \bibinfo {author} {\bibfnamefont {J.}~\bibnamefont {Wang}},
  \bibinfo {author} {\bibfnamefont {A.~A.}\ \bibnamefont {Gentile}}, \bibinfo
  {author} {\bibfnamefont {S.}~\bibnamefont {Paesani}}, \bibinfo {author}
  {\bibfnamefont {N.}~\bibnamefont {Wiebe}}, \bibinfo {author} {\bibfnamefont
  {J.~R.}\ \bibnamefont {McClean}}, \bibinfo {author} {\bibfnamefont
  {S.}~\bibnamefont {Morley-Short}}, \bibinfo {author} {\bibfnamefont {P.~J.}\
  \bibnamefont {Shadbolt}}, \bibinfo {author} {\bibfnamefont {D.}~\bibnamefont
  {Bonneau}}, \bibinfo {author} {\bibfnamefont {J.~W.}\ \bibnamefont
  {Silverstone}}, \bibinfo {author} {\bibfnamefont {D.~P.}\ \bibnamefont
  {Tew}}, \bibinfo {author} {\bibfnamefont {X.}~\bibnamefont {Zhou}}, \bibinfo
  {author} {\bibfnamefont {J.~L.}\ \bibnamefont {O'Brien}},\ and\ \bibinfo
  {author} {\bibfnamefont {M.~G.}\ \bibnamefont {Thompson}},\ }\href
  {https://doi.org/10.1126/sciadv.aap9646} {\bibfield  {journal} {\bibinfo
  {journal} {Science Advances}\ }\textbf {\bibinfo {volume} {4}},\ \bibinfo
  {pages} {eaap9646} (\bibinfo {year} {2018})}\BibitemShut {NoStop}%
\bibitem [{\citenamefont {Zhang}\ \emph {et~al.}(2020)\citenamefont {Zhang},
  \citenamefont {Yuan},\ and\ \citenamefont {Yin}}]{varianceVQE}%
  \BibitemOpen
  \bibfield  {author} {\bibinfo {author} {\bibfnamefont {D.-B.}\ \bibnamefont
  {Zhang}}, \bibinfo {author} {\bibfnamefont {Z.-H.}\ \bibnamefont {Yuan}},\
  and\ \bibinfo {author} {\bibfnamefont {T.}~\bibnamefont {Yin}},\ }\href@noop
  {} {} (\bibinfo {year} {2020}),\ \Eprint {https://arxiv.org/abs/2006.15781}
  {arXiv:2006.15781 [quant-ph]} \BibitemShut {NoStop}%
\bibitem [{\citenamefont {Tang}\ \emph {et~al.}(2020)\citenamefont {Tang},
  \citenamefont {Shkolnikov}, \citenamefont {Barron}, \citenamefont {Grimsley},
  \citenamefont {Mayhall}, \citenamefont {Barnes},\ and\ \citenamefont
  {Economou}}]{qubit-ADAPT}%
  \BibitemOpen
  \bibfield  {author} {\bibinfo {author} {\bibfnamefont {H.~L.}\ \bibnamefont
  {Tang}}, \bibinfo {author} {\bibfnamefont {V.~O.}\ \bibnamefont
  {Shkolnikov}}, \bibinfo {author} {\bibfnamefont {G.~S.}\ \bibnamefont
  {Barron}}, \bibinfo {author} {\bibfnamefont {H.~R.}\ \bibnamefont
  {Grimsley}}, \bibinfo {author} {\bibfnamefont {N.~J.}\ \bibnamefont
  {Mayhall}}, \bibinfo {author} {\bibfnamefont {E.}~\bibnamefont {Barnes}},\
  and\ \bibinfo {author} {\bibfnamefont {S.~E.}\ \bibnamefont {Economou}},\
  }\href@noop {} {} (\bibinfo {year} {2020}),\ \Eprint
  {https://arxiv.org/abs/1911.10205} {arXiv:1911.10205 [quant-ph]} \BibitemShut
  {NoStop}%
\bibitem [{\citenamefont {Claudino}\ \emph {et~al.}(2020)\citenamefont
  {Claudino}, \citenamefont {Wright}, \citenamefont {McCaskey},\ and\
  \citenamefont {Humble}}]{claudinoBenchmarkingAdaptiveVariational2020}%
  \BibitemOpen
  \bibfield  {author} {\bibinfo {author} {\bibfnamefont {D.}~\bibnamefont
  {Claudino}}, \bibinfo {author} {\bibfnamefont {J.}~\bibnamefont {Wright}},
  \bibinfo {author} {\bibfnamefont {A.~J.}\ \bibnamefont {McCaskey}},\ and\
  \bibinfo {author} {\bibfnamefont {T.~S.}\ \bibnamefont {Humble}},\ }\href
  {https://doi.org/10.3389/fchem.2020.606863} {\bibfield  {journal} {\bibinfo
  {journal} {Front. Chem.}\ }\textbf {\bibinfo {volume} {8}},\ \bibinfo {pages}
  {606863} (\bibinfo {year} {2020})}\BibitemShut {NoStop}%
\bibitem [{\citenamefont {Ryabinkin}\ \emph {et~al.}(2020)\citenamefont
  {Ryabinkin}, \citenamefont {Lang}, \citenamefont {Genin},\ and\ \citenamefont
  {Izmaylov}}]{Ryabinkin2020}%
  \BibitemOpen
  \bibfield  {author} {\bibinfo {author} {\bibfnamefont {I.~G.}\ \bibnamefont
  {Ryabinkin}}, \bibinfo {author} {\bibfnamefont {R.~A.}\ \bibnamefont {Lang}},
  \bibinfo {author} {\bibfnamefont {S.~N.}\ \bibnamefont {Genin}},\ and\
  \bibinfo {author} {\bibfnamefont {A.~F.}\ \bibnamefont {Izmaylov}},\ }\href
  {https://pubs.acs.org/doi/10.1021/acs.jctc.9b01084} {\bibfield  {journal}
  {\bibinfo  {journal} {J. Chem. Theory Comput.}\ }\textbf {\bibinfo {volume}
  {16}},\ \bibinfo {pages} {1055} (\bibinfo {year} {2020})}\BibitemShut
  {NoStop}%
\bibitem [{\citenamefont {Gomes}\ \emph {et~al.}(2021)\citenamefont {Gomes},
  \citenamefont {Mukherjee}, \citenamefont {Zhang}, \citenamefont {Iadecola},
  \citenamefont {Wang}, \citenamefont {Ho}, \citenamefont {Orth},\ and\
  \citenamefont {Yao}}]{AVQITE}%
  \BibitemOpen
  \bibfield  {author} {\bibinfo {author} {\bibfnamefont {N.}~\bibnamefont
  {Gomes}}, \bibinfo {author} {\bibfnamefont {A.}~\bibnamefont {Mukherjee}},
  \bibinfo {author} {\bibfnamefont {F.}~\bibnamefont {Zhang}}, \bibinfo
  {author} {\bibfnamefont {T.}~\bibnamefont {Iadecola}}, \bibinfo {author}
  {\bibfnamefont {C.-Z.}\ \bibnamefont {Wang}}, \bibinfo {author}
  {\bibfnamefont {K.-M.}\ \bibnamefont {Ho}}, \bibinfo {author} {\bibfnamefont
  {P.~P.}\ \bibnamefont {Orth}},\ and\ \bibinfo {author} {\bibfnamefont
  {Y.-X.}\ \bibnamefont {Yao}},\ }\href@noop {} {} (\bibinfo {year} {2021}),\
  \Eprint {https://arxiv.org/abs/2102.01544} {arXiv:2102.01544
  [physics.chem-ph]} \BibitemShut {NoStop}%
\bibitem [{\citenamefont {Yao}\ \emph {et~al.}(2021)\citenamefont {Yao},
  \citenamefont {Gomes}, \citenamefont {Zhang}, \citenamefont {Wang},
  \citenamefont {Ho}, \citenamefont {Iadecola},\ and\ \citenamefont
  {Orth}}]{AVQDS}%
  \BibitemOpen
  \bibfield  {author} {\bibinfo {author} {\bibfnamefont {Y.-X.}\ \bibnamefont
  {Yao}}, \bibinfo {author} {\bibfnamefont {N.}~\bibnamefont {Gomes}}, \bibinfo
  {author} {\bibfnamefont {F.}~\bibnamefont {Zhang}}, \bibinfo {author}
  {\bibfnamefont {C.-Z.}\ \bibnamefont {Wang}}, \bibinfo {author}
  {\bibfnamefont {K.-M.}\ \bibnamefont {Ho}}, \bibinfo {author} {\bibfnamefont
  {T.}~\bibnamefont {Iadecola}},\ and\ \bibinfo {author} {\bibfnamefont
  {P.~P.}\ \bibnamefont {Orth}},\ }\href
  {https://doi.org/10.1103/PRXQuantum.2.030307} {\bibfield  {journal} {\bibinfo
   {journal} {PRX Quantum}\ }\textbf {\bibinfo {volume} {2}},\ \bibinfo {pages}
  {030307} (\bibinfo {year} {2021})}\BibitemShut {NoStop}%
\bibitem [{\citenamefont {Kim}\ and\ \citenamefont {Huse}(2013)}]{Kim13}%
  \BibitemOpen
  \bibfield  {author} {\bibinfo {author} {\bibfnamefont {H.}~\bibnamefont
  {Kim}}\ and\ \bibinfo {author} {\bibfnamefont {D.~A.}\ \bibnamefont {Huse}},\
  }\href {https://doi.org/10.1103/PhysRevLett.111.127205} {\bibfield  {journal}
  {\bibinfo  {journal} {Phys. Rev. Lett.}\ }\textbf {\bibinfo {volume} {111}},\
  \bibinfo {pages} {127205} (\bibinfo {year} {2013})}\BibitemShut {NoStop}%
\bibitem [{\citenamefont {Kim}\ \emph {et~al.}(2014)\citenamefont {Kim},
  \citenamefont {Ikeda},\ and\ \citenamefont {Huse}}]{Kim14}%
  \BibitemOpen
  \bibfield  {author} {\bibinfo {author} {\bibfnamefont {H.}~\bibnamefont
  {Kim}}, \bibinfo {author} {\bibfnamefont {T.~N.}\ \bibnamefont {Ikeda}},\
  and\ \bibinfo {author} {\bibfnamefont {D.~A.}\ \bibnamefont {Huse}},\ }\href
  {https://doi.org/10.1103/PhysRevE.90.052105} {\bibfield  {journal} {\bibinfo
  {journal} {Phys. Rev. E}\ }\textbf {\bibinfo {volume} {90}},\ \bibinfo
  {pages} {052105} (\bibinfo {year} {2014})}\BibitemShut {NoStop}%
\bibitem [{\citenamefont {Ryabinkin}\ \emph
  {et~al.}(2018{\natexlab{a}})\citenamefont {Ryabinkin}, \citenamefont
  {Genin},\ and\ \citenamefont {Izmaylov}}]{Ryabinkin2018a}%
  \BibitemOpen
  \bibfield  {author} {\bibinfo {author} {\bibfnamefont {I.~G.}\ \bibnamefont
  {Ryabinkin}}, \bibinfo {author} {\bibfnamefont {S.~N.}\ \bibnamefont
  {Genin}},\ and\ \bibinfo {author} {\bibfnamefont {A.~F.}\ \bibnamefont
  {Izmaylov}},\ }\href {https://pubs.acs.org/doi/abs/10.1021/acs.jctc.8b00943}
  {\bibfield  {journal} {\bibinfo  {journal} {J. Chem. Theory Comput.}\
  }\textbf {\bibinfo {volume} {15}},\ \bibinfo {pages} {249} (\bibinfo {year}
  {2018}{\natexlab{a}})}\BibitemShut {NoStop}%
\bibitem [{\citenamefont {Ryabinkin}\ \emph
  {et~al.}(2018{\natexlab{b}})\citenamefont {Ryabinkin}, \citenamefont
  {Genin},\ and\ \citenamefont {Izmaylov}}]{Ryabinkin2018b}%
  \BibitemOpen
  \bibfield  {author} {\bibinfo {author} {\bibfnamefont {I.~G.}\ \bibnamefont
  {Ryabinkin}}, \bibinfo {author} {\bibfnamefont {S.~N.}\ \bibnamefont
  {Genin}},\ and\ \bibinfo {author} {\bibfnamefont {A.~F.}\ \bibnamefont
  {Izmaylov}},\ }\href {https://aip.scitation.org/doi/10.1063/1.5055357}
  {\bibfield  {journal} {\bibinfo  {journal} {J. Chem. Phys.}\ }\textbf
  {\bibinfo {volume} {149}},\ \bibinfo {pages} {214105} (\bibinfo {year}
  {2018}{\natexlab{b}})}\BibitemShut {NoStop}%
\bibitem [{\citenamefont {Nielsen}\ and\ \citenamefont
  {Chuang}(2002)}]{Nielsen02}%
  \BibitemOpen
  \bibfield  {author} {\bibinfo {author} {\bibfnamefont {M.~A.}\ \bibnamefont
  {Nielsen}}\ and\ \bibinfo {author} {\bibfnamefont {I.}~\bibnamefont
  {Chuang}},\ }\href@noop {} {\emph {\bibinfo {title} {Quantum computation and
  quantum information}}}\ (\bibinfo  {publisher} {Cambridge University Press},\
  \bibinfo {address} {New York},\ \bibinfo {year} {2002})\BibitemShut {NoStop}%
\bibitem [{\citenamefont {Wiersema}\ \emph {et~al.}(2020)\citenamefont
  {Wiersema}, \citenamefont {Zhou}, \citenamefont {de~Sereville}, \citenamefont
  {Carrasquilla}, \citenamefont {Kim},\ and\ \citenamefont
  {Yuen}}]{Wiersema20}%
  \BibitemOpen
  \bibfield  {author} {\bibinfo {author} {\bibfnamefont {R.}~\bibnamefont
  {Wiersema}}, \bibinfo {author} {\bibfnamefont {C.}~\bibnamefont {Zhou}},
  \bibinfo {author} {\bibfnamefont {Y.}~\bibnamefont {de~Sereville}}, \bibinfo
  {author} {\bibfnamefont {J.~F.}\ \bibnamefont {Carrasquilla}}, \bibinfo
  {author} {\bibfnamefont {Y.~B.}\ \bibnamefont {Kim}},\ and\ \bibinfo {author}
  {\bibfnamefont {H.}~\bibnamefont {Yuen}},\ }\href
  {https://doi.org/10.1103/PRXQuantum.1.020319} {\bibfield  {journal} {\bibinfo
   {journal} {PRX Quantum}\ }\textbf {\bibinfo {volume} {1}},\ \bibinfo {pages}
  {020319} (\bibinfo {year} {2020})}\BibitemShut {NoStop}%
\bibitem [{\citenamefont {Abanin}\ \emph {et~al.}(2019)\citenamefont {Abanin},
  \citenamefont {Altman}, \citenamefont {Bloch},\ and\ \citenamefont
  {Serbyn}}]{Abanin2019}%
  \BibitemOpen
  \bibfield  {author} {\bibinfo {author} {\bibfnamefont {D.~A.}\ \bibnamefont
  {Abanin}}, \bibinfo {author} {\bibfnamefont {E.}~\bibnamefont {Altman}},
  \bibinfo {author} {\bibfnamefont {I.}~\bibnamefont {Bloch}},\ and\ \bibinfo
  {author} {\bibfnamefont {M.}~\bibnamefont {Serbyn}},\ }\href
  {https://doi.org/10.1103/revmodphys.91.021001} {\bibfield  {journal}
  {\bibinfo  {journal} {Rev. Mod. Phys.}\ }\textbf {\bibinfo {volume} {91}}
  (\bibinfo {year} {2019})}\BibitemShut {NoStop}%
\bibitem [{\citenamefont {Serbyn}\ \emph {et~al.}(2013)\citenamefont {Serbyn},
  \citenamefont {Papi\ifmmode~\acute{c}\else \'{c}\fi{}},\ and\ \citenamefont
  {Abanin}}]{Serbyn13}%
  \BibitemOpen
  \bibfield  {author} {\bibinfo {author} {\bibfnamefont {M.}~\bibnamefont
  {Serbyn}}, \bibinfo {author} {\bibfnamefont {Z.}~\bibnamefont
  {Papi\ifmmode~\acute{c}\else \'{c}\fi{}}},\ and\ \bibinfo {author}
  {\bibfnamefont {D.~A.}\ \bibnamefont {Abanin}},\ }\href
  {https://doi.org/10.1103/PhysRevLett.111.127201} {\bibfield  {journal}
  {\bibinfo  {journal} {Phys. Rev. Lett.}\ }\textbf {\bibinfo {volume} {111}},\
  \bibinfo {pages} {127201} (\bibinfo {year} {2013})}\BibitemShut {NoStop}%
\bibitem [{\citenamefont {Bauer}\ and\ \citenamefont {Nayak}(2013)}]{Bauer13}%
  \BibitemOpen
  \bibfield  {author} {\bibinfo {author} {\bibfnamefont {B.}~\bibnamefont
  {Bauer}}\ and\ \bibinfo {author} {\bibfnamefont {C.}~\bibnamefont {Nayak}},\
  }\href {https://doi.org/10.1088/1742-5468/2013/09/p09005} {\bibfield
  {journal} {\bibinfo  {journal} {J. Stat. Mech.: Theory Exp.}\ }\textbf
  {\bibinfo {volume} {2013}}\bibinfo  {number} { (09)},\ \bibinfo {pages}
  {P09005}}\BibitemShut {NoStop}%
\bibitem [{\citenamefont {Huse}\ \emph {et~al.}(2014)\citenamefont {Huse},
  \citenamefont {Nandkishore},\ and\ \citenamefont
  {Oganesyan}}]{Nandkishore14}%
  \BibitemOpen
\bibfield  {number} {  }\bibfield  {author} {\bibinfo {author} {\bibfnamefont
  {D.~A.}\ \bibnamefont {Huse}}, \bibinfo {author} {\bibfnamefont
  {R.}~\bibnamefont {Nandkishore}},\ and\ \bibinfo {author} {\bibfnamefont
  {V.}~\bibnamefont {Oganesyan}},\ }\href
  {https://doi.org/10.1103/PhysRevB.90.174202} {\bibfield  {journal} {\bibinfo
  {journal} {Phys. Rev. B}\ }\textbf {\bibinfo {volume} {90}},\ \bibinfo
  {pages} {174202} (\bibinfo {year} {2014})}\BibitemShut {NoStop}%
\bibitem [{\citenamefont {Khemani}\ \emph {et~al.}(2016)\citenamefont
  {Khemani}, \citenamefont {Pollmann},\ and\ \citenamefont
  {Sondhi}}]{Khemani16}%
  \BibitemOpen
  \bibfield  {author} {\bibinfo {author} {\bibfnamefont {V.}~\bibnamefont
  {Khemani}}, \bibinfo {author} {\bibfnamefont {F.}~\bibnamefont {Pollmann}},\
  and\ \bibinfo {author} {\bibfnamefont {S.~L.}\ \bibnamefont {Sondhi}},\
  }\href {https://doi.org/10.1103/PhysRevLett.116.247204} {\bibfield  {journal}
  {\bibinfo  {journal} {Phys. Rev. Lett.}\ }\textbf {\bibinfo {volume} {116}},\
  \bibinfo {pages} {247204} (\bibinfo {year} {2016})}\BibitemShut {NoStop}%
\bibitem [{\citenamefont {Pollmann}\ \emph {et~al.}(2016)\citenamefont
  {Pollmann}, \citenamefont {Khemani}, \citenamefont {Cirac},\ and\
  \citenamefont {Sondhi}}]{Pollmann16}%
  \BibitemOpen
  \bibfield  {author} {\bibinfo {author} {\bibfnamefont {F.}~\bibnamefont
  {Pollmann}}, \bibinfo {author} {\bibfnamefont {V.}~\bibnamefont {Khemani}},
  \bibinfo {author} {\bibfnamefont {J.~I.}\ \bibnamefont {Cirac}},\ and\
  \bibinfo {author} {\bibfnamefont {S.~L.}\ \bibnamefont {Sondhi}},\ }\href
  {https://doi.org/10.1103/PhysRevB.94.041116} {\bibfield  {journal} {\bibinfo
  {journal} {Phys. Rev. B}\ }\textbf {\bibinfo {volume} {94}},\ \bibinfo
  {pages} {041116} (\bibinfo {year} {2016})}\BibitemShut {NoStop}%
\bibitem [{\citenamefont {Pezz{\'{e}}}\ and\ \citenamefont
  {Smerzi}(2009)}]{Pezz2009}%
  \BibitemOpen
  \bibfield  {author} {\bibinfo {author} {\bibfnamefont {L.}~\bibnamefont
  {Pezz{\'{e}}}}\ and\ \bibinfo {author} {\bibfnamefont {A.}~\bibnamefont
  {Smerzi}},\ }\href {https://doi.org/10.1103/physrevlett.102.100401}
  {\bibfield  {journal} {\bibinfo  {journal} {Phys. Rev. Lett.}\ }\textbf
  {\bibinfo {volume} {102}},\ \bibinfo {pages} {100401} (\bibinfo {year}
  {2009})}\BibitemShut {NoStop}%
\bibitem [{\citenamefont {Zhang}\ \emph
  {et~al.}(2021{\natexlab{b}})\citenamefont {Zhang}, \citenamefont {Gomes},
  \citenamefont {Yao}, \citenamefont {Orth},\ and\ \citenamefont
  {Iadecola}}]{code}%
  \BibitemOpen
  \bibfield  {author} {\bibinfo {author} {\bibfnamefont {F.}~\bibnamefont
  {Zhang}}, \bibinfo {author} {\bibfnamefont {N.}~\bibnamefont {Gomes}},
  \bibinfo {author} {\bibfnamefont {Y.}~\bibnamefont {Yao}}, \bibinfo {author}
  {\bibfnamefont {P.~P.}\ \bibnamefont {Orth}},\ and\ \bibinfo {author}
  {\bibfnamefont {T.}~\bibnamefont {Iadecola}},\ }\href
  {https://doi.org/10.6084/m9.figshare.14916282.v3} {\bibinfo {title} {Adaptive
  variational quantum eigensolvers for highly excited states}} (\bibinfo {year}
  {2021}{\natexlab{b}})\BibitemShut {NoStop}%
\end{thebibliography}%

\end{document}